\documentclass[sigconf, screen]{acmart}

\usepackage{graphicx}
\usepackage{algorithm2e}
\usepackage{listings}
\usepackage{hyperref}
\usepackage{tikz}
\usepackage{xspace}
\usepackage{booktabs}
\usepackage{multicol}
\usepackage{multirow}
\usepackage{array}
\usepackage{soul}
\usepackage{subcaption}
\usepackage{enumitem}
\usepackage{amsmath,amsfonts, amsthm} 
\usepackage{xcolor}
\usepackage{pifont}             
\usepackage{balance}
\usepackage[most]{tcolorbox}

\usepackage{multirow}
\usepackage{colortbl}
\usepackage{xcolor}
\begin{document}

\title{CryptRISC: A Secure RISC-V Processor for High-Performance Cryptography with Power Side-Channel Protection}


\author{Amisha Srivastava}
\affiliation{
  \institution{Department of Electrical and Computer Engineering, University of Texas at Dallas}
  \country{USA}
}

\author{Muskan Porwal}
\affiliation{
  \institution{Department of Electrical and Computer Engineering, University of Texas at Dallas}
  \country{USA}
}

\author{Kanad Basu}
\affiliation{
  \institution{Department of Electrical, Computer and Systems Engineering, Rensselaer Polytechnic Institute}
  \country{USA}
}

\begin{abstract} 
Cryptographic computations are fundamental to modern computing, ensuring data confidentiality and integrity. However, these operations are highly susceptible to power side-channel attacks, which exploit variations in power consumption to leak sensitive information. Masking is a widely adopted countermeasure for mitigating side-channel vulnerabilities. However, software-based masking techniques often incur substantial performance overhead and introduce implementation complexity. On the other hand, fixed-function hardware masking is inflexible and often unsuitable for a myriad of cryptographic algorithms. In this paper, we present \textbf{CryptRISC}, the first RISC-V-based processor that integrates cryptographic acceleration with hardware-level power side-channel resistance through an ISA-driven operand masking framework. 
Our design extends the CVA6 core with 64-bit RISC-V Scalar Cryptography Extensions and introduces two key microarchitectural components: a Field Detection Layer (FDL) that identifies the dominant algebraic field of each cryptographic instruction, and a Masking Control Unit (MCU) that applies field-aware operand randomization at runtime. This architecture enables dynamic selection of Boolean, affine, or arithmetic masking schemes based on instruction semantics, allowing optimized protection against power side-channel leakage across cryptographic algorithms such as AES, SHA-256, SHA-512, SM3, and SM4. Unlike prior approaches that rely on fixed masking logic or software instrumentation, our method performs operand masking transparently within the execution pipeline, without modifying the instruction encoding. Experimental results demonstrate that CryptRISC achieves speedups up to 6.80$\times$ compared to baseline software implementations while providing robust side-channel resistance. Despite these enhancements, CryptRISC incurs only a 1.86\% increase in hardware overhead over the baseline CVA6 core, confirming its efficiency and practicality. Overall, CryptRISC provides a scalable foundation for integrating cryptographic acceleration with runtime side-channel protection, offering strong security guarantees, minimal overhead, and full ISA compatibility.

\end{abstract}



\maketitle

\section{Introduction}
\label{Introduction}

Cryptographic computations play a crucial role in modern computing, ensuring data confidentiality, integrity, and authentication across various domains, including secure communications, financial transactions, and embedded system security. However, executing encryption and hashing operations on general-purpose processors imposes a significant computational burden due to the reliance on intensive bitwise operations, modular arithmetic, and state transformations \cite{rifa2011computational}. As a result, cryptographic acceleration has become essential, typically implemented using one of three approaches: 1) software-based implementations, 2) dedicated hardware accelerators, or 3) hybrid instruction set extensions.

Software-based cryptographic implementations provided by libraries like OpenSSL, WolfSSL and Libgcrypt offer flexibility but suffer from high execution latency, particularly in constrained environments such as embedded systems \cite{viega2002network, lacy1993cryptolib}. For instance, performing AES encryption in software on a 32-bit processor may require hundreds of cycles per block, making it unsuitable for real-time applications. Dedicated hardware accelerators, such as OpenTitan AES hardware IP, address this performance bottleneck by offloading cryptographic computations to specialized circuitry \cite{ciani2024unleashing}. While these accelerators significantly reduce execution time, they are often fixed-function, making it difficult to integrate evolving security countermeasures or adapt to new cryptographic standards. Instruction set extensions (ISEs) provide a balanced solution by incorporating specialized cryptographic instructions directly into general-purpose processors, enabling efficient execution of cryptographic operations within the processor pipeline while maintaining programmability and reducing software overhead \cite{galuzzi2011instruction}.

\begin{figure}[t!] \centering \includegraphics[width=\linewidth]{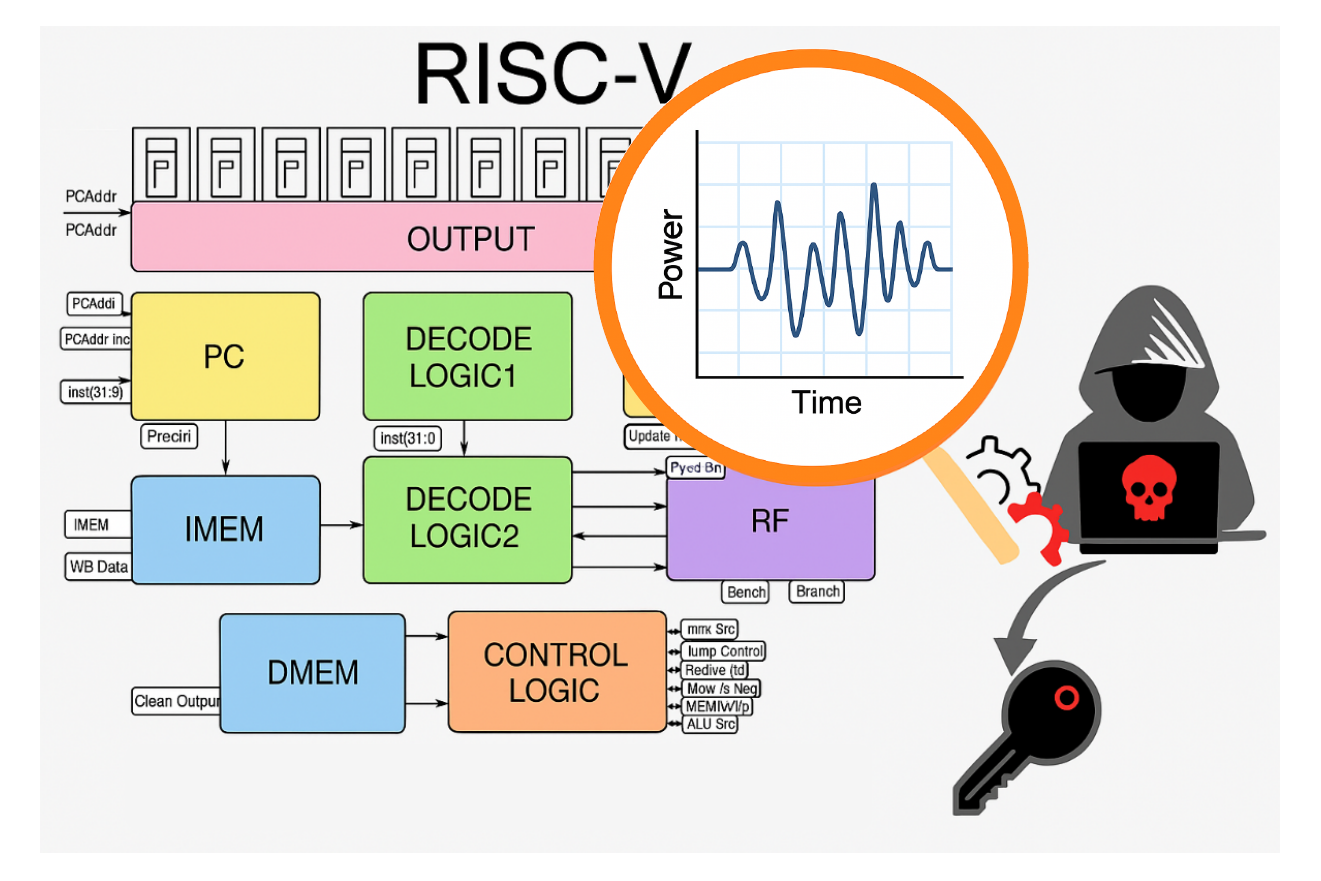} \caption{Power Side-Channel Attack flow on a RISC-V Processor: Attacker can exploit the power consumption variations during cryptographic operations to reveal sensitive data by analyzing power leakage at different stages of instruction execution of the processor.} \label{fig:intro_psca} \end{figure}

RISC-V has emerged as a widely adopted, flexible, and open Instruction Set Architecture (ISA) known for its extensibility and adaptability across various computing domains, from deeply embedded IoT devices to high-performance server-class machines. The availability of custom ISEs has facilitated domain-specific optimizations, particularly for cryptographic acceleration. While these extensions improve performance, they were primarily designed with efficiency in mind, and security, especially in the context of side-channel attacks, was not a central concern. Traditional cryptographic systems often rely on the two extremes: software-based cryptography, which is flexible but slow, and dedicated hardware accelerators, which are fast but lack adaptability. Cryptographic ISEs provide a middle ground by integrating cryptographic acceleration into general-purpose processors, but they still lack integrated countermeasures against power side-channel vulnerabilities \cite{bartolini2009instruction, marshall2021design}.

Power side-channel attacks, such as Differential Power Analysis and Correlation Power Analysis, exploit variations in power consumption during cryptographic computations to extract secret keys \cite{oswald2005efficient}. These attacks are particularly effective against hardware implementations, as predictable execution patterns can reveal sensitive information through power leakage, as illustrated in Figure \ref{fig:intro_psca}. Existing research has shown that RISC-V cryptographic instruction set
extension (CISE) based designs, such as XCRYPTO \cite{xcrypto} and RISCV-CRYPTO \cite{riscvcrypto}, excel in accelerating cryptographic operations. However, they remain vulnerable to these power side-channel attacks due to the lack of integrated countermeasures \cite{jayasena2025ciseleaks}. Cryptographic instructions, which manipulate secret-dependent data, may leak information through fluctuations in power consumption, making them susceptible to such attacks.\textcolor{black}{While shrinking technology nodes reduce the absolute magnitude of power signals, they also increase signal-to-noise ratio and sensor resolution, making it easier for sophisticated adversaries to extract fine-grained power signatures from modern chips.} This highlights the pressing need to enhance RISC-V's cryptographic instruction set with built-in resistance to power side-channel vulnerabilities, especially as RISC-V gains traction in security-critical applications.

To mitigate power side-channel vulnerabilities, various hardware-based countermeasures have been proposed, including hiding and masking techniques. Hiding techniques aim to reduce power leakage correlation by introducing execution randomness, such as instruction shuffling or clock jittering \cite{fritzke2012obfuscating}. However, these approaches require complex control logic and can still be susceptible to advanced statistical attacks. Masking techniques, on the other hand, introduce randomization at the data level by encoding cryptographic variables with random masks, preventing direct leakage of secret key information \cite{messerges2000securing, damgaard2010perfectly}. While software-based masking offers flexibility, it introduces significant computational and memory overhead \cite{moriai2014fast}. Fixed-function hardware masking, while reducing overhead, lacks adaptability across different cryptographic algorithms \cite{uhleanother}. A major limitation in current RISC-V cryptographic extensions is their reliance on fixed or software-based masking, which is inefficient because different cryptographic algorithms require distinct masking techniques. For example, AES encryption and decryption benefit from Boolean masking due to their XOR-based structure, SHA-256 and SHA-512 require affine masking to protect bitwise logical operations, and SM3 and SM4 involve modular arithmetic, making arithmetic masking the most effective countermeasure \cite{ji2009algebraic, martinkauppi2020design}. \textit{A one-size-fits-all masking approach is suboptimal, as it introduces unnecessary computational costs while leaving certain algorithms vulnerable.}

To address the growing threat of power side-channel attacks in cryptographic hardware, we present \textbf{CryptRISC}, the first RISC-V processor that combines cryptographic acceleration with integrated, field-aware masking support through a novel ISA-driven operand randomization mechanism. Our design extends the open-source CVA6 core with 64-bit RISC-V Scalar Cryptography Extensions and introduces a systematic hardware-level countermeasure that dynamically applies masking transformations based on the underlying algebraic field of each instruction. This is achieved through two architectural features: a lightweight Field Detection Layer (FDL) embedded in the decode stage, which identifies the dominant algebraic domain of each cryptographic instruction, and a Masking Control Unit (MCU) integrated into the execution stage, which securely transforms operands using field-appropriate masking schemes.

Unlike prior approaches that rely on software-inserted masking logic or fixed datapath-level countermeasures, CryptRISC achieves flexible and efficient side-channel resistance by dynamically selecting masking strategies—Boolean, affine, or arithmetic—based on instruction semantics. By embedding this logic directly into the pipeline, CryptRISC avoids instruction re-encoding and software overhead, while offering scalable protection tailored to the cryptographic workload.

The main contributions of this paper are as follows:

\begin{itemize} \item We introduce CryptRISC, a RISC-V-based processor enhanced with cryptographic ISE and a novel hardware-based instruction-level masking architecture that dynamically selects masking strategies based on field identification at runtime. 

\item The masking mechanism is fully integrated into the hardware pipeline through the Field Detection Layer (FDL) and Masking Control Unit (MCU), requiring no changes to the instruction encoding. This enables transparent operand randomization without any software modifications.

\item By employing masking strategies to align with the algebraic field characteristics of each instruction, CryptRISC provides robust protection against PSC attacks across diverse cryptographic workloads. This field-aware operand masking is validated through instruction-level Test Vector Leakage Assessment (TVLA), with the majority of observed $t$-values falling within $\pm2$, and all values remaining well below the leakage threshold of $\pm4.5$, indicating the absence of statistically significant first-order leakage.

\item We evaluated the performance of CryptRISC using the RISC-V cryptographic benchmarks suite, which includes algorithms such as AES, SHA-256, SHA-512, SM3, and SM4. To enable efficient hardware acceleration, we integrated the RISC-V Scalar Cryptography Extension into the CVA6 core. Our results demonstrate that this enhancement significantly improves cryptographic performance, achieving speedups of up to 6.80$\times$ over baseline software implementations.


\end{itemize}

\section{Background}

\subsection{Overview of Algebraic Fields}
\label{subsubsec:fields_background}

Cryptographic algorithms operate within specific algebraic structures known as fields, which define the rules for arithmetic and logical operations \cite{murty2010algebraic}. A field is a set equipped with two binary operations—addition and multiplication—that satisfy a complete set of axioms, including associativity, commutativity, distributivity, and the existence of additive and multiplicative inverses \cite{nicholson2012introduction}. Fields enable the construction of rich algebraic systems where both linear and non-linear transformations can be reliably performed and reversed \cite{shparlinski2013finite}.
In cryptography, we primarily encounter \textbf{finite fields}, or \emph{Galois Fields}, denoted as \(GF(p^n)\), where \(p\) is a prime number and \(n\) is a positive integer \cite{benvenuto2012galois}. These fields contain a finite number of elements (\(p^n\)) and form the basis for efficient implementations of block ciphers, hash functions, and modular arithmetic. Three primary classes of algebraic fields play a critical role in cryptography:
\begin{itemize}
    \item \textbf{Binary Fields} (\(GF(2^n)\)) are used extensively in symmetric-key ciphers such as AES and SM4. Elements in these fields are \(n\)-bit vectors, and arithmetic operations are performed modulo an irreducible polynomial. These fields enable efficient implementation of non-linear S-boxes and matrix-based linear transformations.
    
    \item \textbf{Boolean Algebra} is a degenerate case of \(GF(2)\), where operations reduce to bitwise logic (e.g., \texttt{XOR}, \texttt{AND}, rotations). It forms the backbone of hash functions like SHA-2 and SM3, where computations are expressed in binary terms.
    
    \item \textbf{Modular Integer Rings} (\(\mathbb{Z}/2^n\mathbb{Z}\)) are not technically fields due to the lack of multiplicative inverses for all non-zero elements. However, they are ubiquitous in hash function designs that require word-level modular additions, such as those found in SHA-256 or SM3.
\end{itemize}

\subsection{Power Side-Channel Attacks}

Power Side Channel (PSC) attacks are techniques that leverage variations in a device's power consumption to undermine cryptographic security \cite{randolph2020power}. These attacks are categorized into three primary types: Simple Power Analysis (SPA), Differential Power Analysis (DPA), and Correlation Power Analysis (CPA).
SPA involves the direct observation of power consumption patterns during the execution of cryptographic algorithms \cite{mangard2003simple}. By analyzing these patterns, an attacker can infer critical data and understand how it interacts with the algorithm during processing. DPA 
is a more sophisticated technique that involves the statistical analysis of power consumption data collected over multiple encryption or decryption operations \cite{kocher1999differential, kocher2011introduction}. CPA is a further refinement of DPA. It uses a more advanced statistical method to analyze the correlation between power consumption and the intermediate values within cryptographic computations \cite{brier2004correlation}. Therefore, the effectiveness of these attacks underscores the necessity for early detection and mitigation of potential vulnerabilities in cryptographic hardware designs \cite{srivastava2024scar, zhang2021psc, he2019rtl}.

\subsection{RISC-V Scalar Cryptography Extensions}

The RISC-V architecture has incorporated Scalar Cryptography Extensions to enhance the efficiency of symmetric cryptographic operations \cite{zeh2021risc}. These extensions introduce specialized instructions for algorithms such as AES, SHA-256, SHA-512, SM3, and SM4, significantly reducing execution cycles compared to software implementations. The Scalar Cryptography Extensions consist of the specialist cryptographic instructions that are dedicated execution units to handle specific operations for cryptographic algorithms. The formal release of these extensions, known as Scalar Cryptography v1.0.1, facilitates the development of secure and efficient cryptographic applications on RISC-V processors \cite{spec}.\textcolor{black}{These instructions aim to boost cryptographic performance, not to address physical side-channel resilience. While the specification mandates timing side-channel mitigation, power and EM leakage defenses are explicitly left to microarchitectural or physical implementations~\cite{spec}. Consequently, processors remain vulnerable to PSC attacks without dedicated hardware protections.}



\begin{table*}[t!]
\caption{Representative 64-bit RISC-V Scalar Cryptographic Instructions}
\label{tab:riscv_scalar_instructions}
\centering
\begin{tabular}{l l p{6cm}}
    \toprule
    \textbf{Instruction} & \textbf{Algorithm} & \textbf{Purpose} \\
    \midrule
    \texttt{saes64.encs}, \texttt{saes64.encsm}, \texttt{saes64.ks1}, \texttt{saes64.ks2}, \texttt{saes64.im} &
    AES &
    Encryption round steps, key schedule, Inverse MixColumns \\
    
    \texttt{saes64.ds}, \texttt{saes64.dsm} &
    AES &
    Decryption round steps, masked decrypti enon \\
    
    \texttt{sm4.ed}, \texttt{sm4.ks} &
    SM4 &
    Block encryption/decryption, key scheduling \\
    
    \texttt{ssm3.p0}, \texttt{ssm3.p1} &
    SM3 &
    Permutation steps for SM3 hashing \\
    
    \texttt{ssha256.sig0}, \texttt{ssha256.sig1}, \texttt{ssha256.sum0}, \texttt{ssha256.sum1} &
    SHA-256 &
    Logical $\oplus$ and rotate-based round operations \\
    
    \texttt{ssha512.sig0}, \texttt{ssha512.sig1}, \texttt{ssha512.sum0}, \texttt{ssha512.sum1} &
    SHA-512 &
    64-bit variants of rotation-based round transformations \\
    \bottomrule
\end{tabular}
\end{table*}

\begingroup
\color{black}
\section{Related Works} \label{sec:motivation}

The adoption of RISC-V cryptographic extensions has renewed focus on mitigating PSC attacks. Traditional countermeasures fall into two main categories: software-based masking and fixed-function hardware masking. Software-based masking modifies cryptographic code to introduce randomness \cite{oswald2005efficient}, but this approach imposes high overhead and is sensitive to coding practices and compiler behavior \cite{moriai2014fast, venkatesh2019security, beckers2022provable}. Instruction Set Extensions (ISEs) have been proposed to alleviate performance bottlenecks \cite{gao2021instruction}, but still require manual integration. Fixed-function hardware masking incorporates Boolean or arithmetic masking schemes directly into dedicated hardware units \cite{sasdrich2020low, rivain2010provably, fritzmann2022masked}. These designs reduce overhead but lack flexibility, as they are not adaptable to different algorithms \cite{uhleanother}.

\endgroup
\section{CryptRISC: Enhancing Cryptographic Acceleration in the CVA6 Core}
\label{subsec:crypto_accel_cva6}

The computational intensity of cryptographic operations—particularly encryption, hashing, and signature generation—often necessitates dedicated hardware support to ensure efficient execution. The CVA6 core, a high-performance, 64-bit RISC-V processor, is well-suited for accelerating cryptographic workloads while maintaining architectural flexibility \cite{cva6}. CVA6 supports the custom RISC-V Scalar Cryptography Extensions, which introduce specialized cryptographic instructions that significantly improve the execution efficiency of symmetric encryption algorithms, hash functions, and other bitwise and modular arithmetic operations commonly used in cryptographic applications. These extensions allow cryptographic workloads to be offloaded to dedicated hardware-optimized instructions, reducing execution latency and power consumption compared to software-based implementations. Table~\ref{tab:riscv_scalar_instructions} lists the key 64-bit instructions included in the ratified RISC-V Scalar Cryptography Extensions. They primarily target algorithms such as \textbf{AES}, \textbf{SM4}, \textbf{SM3}, \textbf{SHA-256}, and \textbf{SHA-512}, focusing on accelerating critical operations like S-box lookups, bitwise logic, and modular addition.


While the CISE designs accelerate cryptographic operations, they offer no inherent protection against PSC attacks. To address this challenge, we introduce \textbf{CryptRISC}, a RISC-V processor based on the CVA6 core, which extends the architecture to include 64-bit RISC-V Scalar Cryptography Extensions. CryptRISC integrates both cryptographic acceleration and field-aware side-channel countermeasures, offering an efficient and secure solution for cryptographic operations. Further details on these enhancements and countermeasures are discussed in the following sections.

\begin{figure*}[t!]
    \centering
    \includegraphics[width=\linewidth]{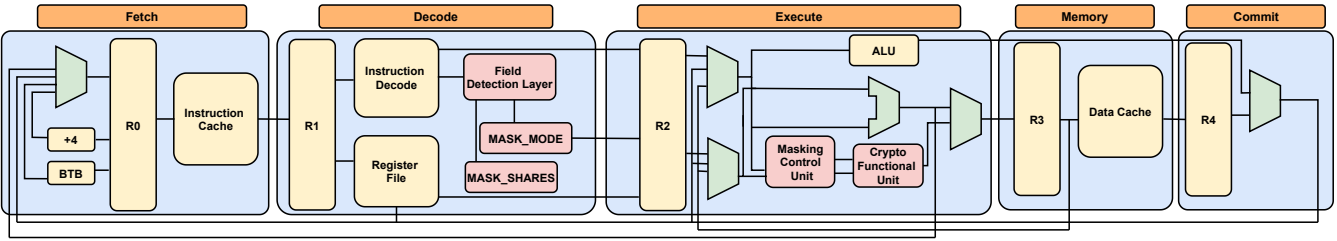}
    \caption{Overview of pipeline flow in the proposed side-channel hardened processor, CryptRISC, with architectural blocks introduced in this work highlighted in red.}
    \label{fig:vulne}
\end{figure*}

\section{Integration of 64-bit RISC-V Scalar Crypto Instructions into CVA6 for CryptRISC}
\label{subsec:scalar_crypto_into_cva6}

To enhance cryptographic performance, \textbf{CryptRISC} extends the CVA6 core with the \textbf{64-bit RISC-V Scalar Cryptography Extensions}, adding specialized instructions for accelerating symmetric encryption, hashing, and modular arithmetic.  To support hardware execution of these instructions, we incorporate a dedicated \textbf{Crypto Functional Unit (CFU)} into the CVA6 pipeline as part of the \textbf{CryptRISC} design. The CFU is added as a new execution unit within the \textit{Execute} stage and is responsible for implementing the cryptographic operations specified by the RISC-V scalar crypto ISA extensions. The key modifications to the pipeline include the following:

\begin{itemize}
    \item \emph{Crypto Functional Unit (CFU):} We introduce a new CFU that performs field-specific operations such as AES S-box substitutions, modular additions, logical rotations, and bitwise logic. The CFU is invoked exclusively for scalar cryptographic opcodes and is tightly integrated into the existing execute stage.
    
    \item \emph{Pipeline Extension:} The six-stage CVA6 pipeline (\textit{Fetch}, \textit{Decode}, \textit{Register Read}, \textit{Execute}, \textit{Memory Access}, \textit{Write-Back}) is modified in \textbf{CryptRISC} to recognize scalar crypto instructions and dispatch them to the CFU via a dedicated datapath.

    \item \emph{Operand Forwarding:} The operand forwarding logic is extended to support data dependencies involving cryptographic instructions, ensuring that results produced by the CFU can be consumed with minimal stalls.

    \item \emph{Memory Subsystem Support:} The integration maintains compatibility with CVA6's memory hierarchy, allowing efficient access to round keys and intermediate states required by long-running hash or block cipher computations.
\end{itemize}

These extensions allow CVA6 to execute cryptographic workloads in a secure and fast manner, providing a hardware foundation for applications involving authenticated encryption, secure hashing, and signature verification. Furthermore, the modular integration of the CFU ensures that the architectural changes remain localized, preserving the pipeline’s flexibility and maintainability.
 
\textcolor{black}{While integrating these instructions resolves performance bottlenecks in cryptographic workloads, it does not mitigate PSC attacks, as hardware-accelerated cryptographic implementations can still leak sensitive information through subtle variations in power consumption \cite{jayasena2025ciseleaks}. These designs follow the standard ISA specification but optimize for performance without incorporating physical or microarchitectural countermeasures, thereby making instruction-level leakage observable via power analysis.} Consequently, in the following section, we introduce a dynamic masking scheme that embeds side-channel countermeasures within CVA6, as part of the CryptRISC design, at the instruction level. This approach minimizes additional overhead while bolstering security. 

\section{Field-Aware Dynamic Masking in CryptRISC}
\label{subsec:field_aware_masking}

This section introduces the central idea behind our field-aware masking architecture for preventing PSC attacks in CryptRISC, as illustrated in Figure \ref{fig:vulne}. Our work presents an ISA-driven dynamic masking framework that selects the most suitable masking scheme at the granularity of individual cryptographic instructions. Central to our design is a \textbf{Field-Detection Layer (FDL)} in the decode stage of the processor pipeline, which analyzes the instruction opcode and identifies the dominant algebraic structure. This field information then informs the selection of an appropriate masking style: Boolean, arithmetic, affine, or multiplicative. To encode this decision within the instruction stream, we extend the RISC-V Scalar Cryptography ISA with both the masking type and its security level. At runtime, this information is interpreted by a dedicated \textbf{Masking Control Unit (MCU)} in the execute stage, which applies the appropriate operand transformation using a reconfigurable affine masking engine. Together, these components form a unified and flexible masking architecture that adapts dynamically to the structure of the cryptographic workload. By leveraging instruction-level field-awareness, generalized affine masking, and lightweight ISA extensions, our system delivers robust protection against power-based attacks in CryptRISC while preserving compatibility and performance across diverse cryptographic algorithms.


\subsection{Dominant Algebraic Fields in Cryptographic Algorithms}
\label{subsubsec:fields_across_algos}

A key consideration in implementing hardware countermeasures for side-channel attacks is understanding the algebraic structure over which cryptographic operations are performed. Each algebraic domain exhibits distinct switching characteristics that influence its power leakage profile. Accordingly, the selection of an appropriate operand masking scheme—whether Boolean, arithmetic, or affine is fundamentally guided by the field in which the computation occurs. Modern cryptographic algorithms often consist of mixed-field operations. However, most are dominated by one or two algebraic structures that contribute disproportionately to side-channel leakage. Identifying these \emph{dominant fields} enables the architecture to apply field-specific masking schemes that achieve optimal leakage reduction with minimal overhead. We analyze the field characteristics of some commonly used algorithms below:

\begin{itemize}
    \item Block ciphers such as AES and SM4 operate primarily over the finite field \(GF(2^8)\), utilizing non-linear substitution layers (S-boxes) and polynomial transformations \cite{ji2009algebraic}. These field operations are highly susceptible to data-dependent power leakage, especially during table lookups and multiplications. As a result, affine and multiplicative masking techniques are preferred for securing these operations.
    \item Hash functions in the SHA-2 family combine logical operations (e.g., XOR, AND, rotations) with modular additions. The former are naturally expressed in \(GF(2)\), while the latter take place in integer rings such as \(\mathbb{Z}/2^{32}\mathbb{Z}\) or \(\mathbb{Z}/2^{64}\mathbb{Z}\) \cite{martinkauppi2020design}. As each domain exhibits different leakage characteristics, SHA-2 requires a hybrid strategy that applies Boolean masking for logic and arithmetic masking for modular additions.
    \item SM3 resembles SHA-256 in structure and leakage profile, involving both logical operations and modular arithmetic \cite{martinkauppi2020design}. Consequently, it benefits from the same hybrid masking strategy to ensure comprehensive side-channel resistance.
\end{itemize}

Therefore, field-aware masking strategies are essential for tailoring countermeasures to the leakage profiles of different cryptographic operations. Recognizing the dominant algebraic domain of each instruction enables hardware to apply the most effective randomization technique without compromising performance.

\subsection{Field Detection Layer} \label{subsubsec:hardware_field_detection}

Field-aware operand masking requires hardware to recognize the underlying algebraic structure of each cryptographic instruction. To enable this capability, we introduce a \textbf{Field Detection Layer (FDL)} into the \texttt{Decode} stage of the CryptRISC pipeline. The FDL statically classifies each instruction based on its semantic role and corresponding leakage domain, allowing the processor to select an appropriate masking scheme during execution.

Each instruction introduced by the RISC-V Scalar Cryptography Extensions is associated with a specific cryptographic primitive—such as AES round functions, SHA-2 mixing operations, or SM3 permutations, and is thus tied to a dominant algebraic field. These dominant fields are identified through the analysis of the algorithm’s structure and leakage characteristics, as discussed in Section~\ref{subsubsec:fields_across_algos}. The FDL uses this information to associate each instruction with a symbolic \textit{field tag} that captures the relevant algebraic domain. \textcolor{black}{This field-tag-based selection process does not depend on runtime data or secret-dependent conditions. It is determined statically at decode time based solely on the instruction opcode, and therefore does not introduce any new side-channel leakage.}

\subsubsection{\textbf{FDL Architecture.}}
The FDL is implemented as a lightweight combinational logic block within the decode stage of the CryptRISC core. Its primary component is a \textit{lookup table} (LUT) indexed by the opcode of scalar cryptographic instructions. Each entry in the LUT maps an instruction to one or more predefined symbolic field tags—such as \textit{FIELD\_GF2}, \textit{FIELD\_GF2N}, or \textit{FIELD\_Z2N}—that represent Boolean logic, finite-field arithmetic, and modular integer arithmetic, respectively. These tag assignments are fixed at design time based on the dominant field classification and are summarized in Table~\ref{tab:field_tag_lookup}. Upon lookup, the FDL passes a symbolic field tag that is forwarded along with the decoded instruction. This tag is used in the next stage to derive fine-grained masking metadata, which are used to drive operand randomization as detailed in Section~\ref{subsubsec:isa_extensions}.

\begin{table}[h]
\centering
\caption{Field Tag Classification for RISC-V Scalar Cryptographic Instructions}
\label{tab:field_tag_lookup}
\begin{tabular}{l p{2.5cm}}
    \toprule
    \textbf{Instruction(s)} & \textbf{Field Tag(s)} \\
    \midrule
    \texttt{saes64.encs}, \texttt{saes64.encsm}, \texttt{saes64.im} & \texttt{FIELD\_GF2N} \\
   \texttt{saes64.ks11}, \texttt{saes64.ks2} & \texttt{FIELD\_GF2N} \\
    \texttt{saes64.ds}, \texttt{saes64.dsm} & \texttt{FIELD\_GF2N} \\
    \texttt{sm4.ed}, \texttt{sm4.ks} & \texttt{FIELD\_GF2N} \\
\texttt{ssm3.p0} & \texttt{FIELD\_GF2} \\
\texttt{ssm3.p1} & \texttt{FIELD\_Z2N} \\
    \texttt{ssha256.sig0}, \texttt{ssha256.sig1} & \texttt{FIELD\_GF2} \\
    \texttt{ssha256.sum0}, \texttt{ssha256.sum1} & \texttt{FIELD\_Z2N} \\
    \texttt{ssha512.sig0}, \texttt{ssha512.sig1} & \texttt{FIELD\_GF2} \\
    \texttt{ssha512.sum0}, \texttt{ssha512.sum1} & \texttt{FIELD\_Z2N} \\
    \bottomrule
\end{tabular}
\end{table}

\subsubsection{\textbf{Extensibility}}
The FDL can be easily extended to support new instruction formats or cryptographic extensions by augmenting its LUT entries. No changes are required to the masking datapath or instruction format, maintaining backward compatibility and design modularity. Therefore, the FDL provides a hardware mechanism for mapping cryptographic instructions to their dominant algebraic domains, enabling per-instruction, field-aware masking. It acts as the frontend to the processor’s operand masking infrastructure, supplying classification metadata that guides the secure execution of cryptographic workloads.

\subsubsection{\textbf{Microarchitectural Masking Configuration}}
\label{subsubsec:isa_extensions}

While the original RISC-V Scalar Cryptography ISA does not provide built-in mechanisms for power side-channel resistance, we introduce a lightweight architectural extension that enables dynamic, instruction-specific operand masking entirely in hardware. Rather than modifying the instruction encoding or ISA format, our approach derives two internal microarchitectural metadata fields—\textit{MASK\_MODE} and \textit{MASK\_SHARES}—within the decode stage using information provided by the Field Detection Layer (FDL).

As described previously, the FDL analyzes each cryptographic instruction’s opcode and emits a symbolic field tag that identifies the dominant algebraic domain (e.g., \(GF(2)\), \(GF(2^n)\), or \(\mathbb{Z}/2^n\mathbb{Z}\)). This symbolic tag is not directly consumed by the masking hardware; instead, it is interpreted by a dedicated mapping unit within the decode stage, which translates the tag into concrete metadata fields that determine the masking policy for the instruction.

\begin{itemize}
    \item \textbf{\texttt{MASK\_MODE}}: Encodes the type of masking transformation to be applied, as derived from the field tag. Table~\ref{tab:mask_mode_encoding} provides a mapping between field tags and their corresponding \texttt{MASK\_MODE} values, along with the masking technique suitable for each field.

    \item \textbf{\texttt{MASK\_SHARES}}: Indicates the number of random shares to generate for operand masking. For example, \texttt{MASK\_SHARES} = 1 enables first-order masking, while higher values support second- or higher-order countermeasures against multi-trace attacks. This value may be determined based on a fixed security policy or a programmable mapping indexed by the field tag.
\end{itemize}


\begin{table}[h]
\centering
\caption{Mapping Between Field Tags and \texttt{MASK\_MODE}}
\label{tab:mask_mode_encoding}
\begin{tabular}{l c l}
    \toprule
    \textbf{Field Tag} & \textbf{MASK\_MODE} & \textbf{Masking Type} \\
    \midrule
    \texttt{FIELD\_GF2}   & \texttt{01} & Boolean masking \\
    \texttt{FIELD\_GF2N}  & \texttt{10} & Affine / Multiplicative masking \\
    \texttt{FIELD\_Z2N}   & \texttt{11} & Arithmetic masking \\
    \midrule
    \texttt{DEFAULT}      & \texttt{00} & No masking \\
    \bottomrule
\end{tabular}
\end{table}

These metadata fields are not encoded in the instruction binary but are instead attached to each instruction internally and propagated alongside it through the pipeline. They serve as control signals to the Masking Control Unit (MCU), which performs operand masking in the \texttt{Execute} stage. Figure~\ref{fig:masking_metadata_mapping} illustrates this tagging and propagation mechanism.

\begin{figure}[h]
\centering
\[
\texttt{Field Tag} \xrightarrow{} \quad
\underbrace{\texttt{MASK\_MODE}}_{2\text{ bits}} \quad + \quad
\underbrace{\texttt{MASK\_SHARES}}_{2\text{ bits}}
\]
\caption{Translation of the symbolic field tag into masking metadata fields during decode.}
\label{fig:masking_metadata_mapping}
\end{figure}

For example, the instruction \texttt{saes64.encs}, which performs an AES encryption step over \(GF(2^8)\). The FDL classifies this as \texttt{FIELD\_GF2N}. The associated mapping logic then sets \textit{MASK\_MODE} to \textit{10} (affine masking), and the policy may assign \textit{MASK\_SHARES} = 2 to enable second-order protection.

The benefits of incorporating FDL are as follows:
\begin{itemize}
    \item \textbf{ISA Compatibility:} No modifications are required to the RISC-V instruction encoding or compiler toolchain.
    \item \textbf{Fine-grained Control:} Masking strategy and security level are customized at the instruction level.
    \item \textbf{Scalable and Flexible:} The configuration logic is fully programmable and easily extended to support various masking styles or threat models.
\end{itemize}

The following section describes how this configuration metadata is consumed by the Masking Control Unit to perform secure, field-aware operand randomization during cryptographic execution.

\subsection{Masking Control Unit}
\label{subsubsec:mcu_affine}

To enable secure and field-aware operand randomization in hardware, we introduce a dedicated \textbf{Masking Control Unit (MCU)} within the \texttt{Execute} stage of the CryptRISC pipeline. This unit applies instruction-specific masking transformations to sensitive operands prior to their consumption by cryptographic functional units, based on metadata derived during decode. The MCU serves as the enforcement point for our dynamic masking architecture, ensuring that all cryptographic computations operate on randomized operand values to suppress data-dependent power leakage.

\subsubsection{\textbf{Microarchitectural Placement and Inputs.}}
The MCU is integrated into the \texttt{Execute} stage, immediately following the operand selection multiplexers. It is placed on the critical data path that connects operand sources (register file or bypass network) to the CFU, ensuring that masking is applied before any sensitive computation occurs.

The MCU receives the following runtime inputs, all forwarded from the \texttt{R2} stage of the pipeline (as shown in Figure \ref{fig:vulne}):
\begin{itemize}
    \item \textbf{Decoded instruction:} Carries opcode-level semantics used for identifying cryptographic operations and associating field-specific constraints.
    \item \textbf{Two unmasked operands:} Fetched from the register file or bypass path, and selected via operand multiplexers. These operands represent sensitive data to be randomized prior to cryptographic execution.
    \item \textbf{\textit{MASK\_MODE}}: A 2-bit microarchitectural field specifying the type of masking to apply (Boolean, affine, or arithmetic), derived from field classification in the decode stage.
    \item \textbf{\textit{MASK\_SHARES}}: A metadata field indicating the number of random shares to generate per operand, enabling runtime configurability for higher-order side-channel resistance.
\end{itemize}

\textcolor{black}{
To prevent data-dependent switching activity, the MCU randomizes operands using fresh affine masks (\(A \cdot x + B\)), ensuring that switching patterns are independent of sensitive values. This disrupts correlation with Hamming weight/distance models commonly exploited in power side-channel attacks.
}

\subsubsection{\textbf{Affine Masking.}}
Modern cryptographic instruction sets span diverse algebraic domains—including \(GF(2)\), \(GF(2^n)\), and \(\mathbb{Z}/2^n\mathbb{Z}\)—each with its own masking requirements. A conventional approach would involve distinct datapaths for Boolean, arithmetic, or multiplicative masking, increasing hardware complexity, control burden, and area overhead. 

To address this challenge, we adopt an \textit{affine masking formulation} that generalizes all common masking strategies under a single, unified transformation \cite{fumaroli2010affine}. This approach is highly expressive and configurable, allowing the same logic to implement multiple masking paradigms based on metadata-guided runtime selection. It enables flexible, per-instruction operand protection without duplicating circuitry or requiring masking-specific hardware blocks.

Affine masking is a generalized form of masking where the MCU transforms each input operand \(x\) into a masked version \(x'\) using the affine function:
\[
x' = A \cdot x + B
\]
where:
\begin{itemize}
    \item \(x\) is the sensitive unmasked operand,
    \item \(A\) is a randomly sampled multiplicative mask,
    \item \(B\) is a randomly sampled additive mask,
    \item All arithmetic is performed over the algebraic domain corresponding to the instruction (e.g., \(GF(2)\), \(GF(2^8)\), or \(\mathbb{Z}/2^n\mathbb{Z}\)).
\end{itemize}
Affine masking serves as the foundation for various masking techniques, with specific configurations of \(A\) and \(B\) giving rise to different masking types:
\begin{itemize}
    \item \textbf{Boolean masking (GF(2))}: \(A = 1\), \(B = r\) yields \(x' = x \oplus r\)
    \item \textbf{Arithmetic masking (\(\mathbb{Z}/2^n\mathbb{Z}\))}: \(A = 1\), \(B = r\) yields \(x' = x + r \mod 2^n\)
    \item \textbf{Multiplicative masking (GF(\(2^n\)))}: \(A = r\), \(B = 0\) yields \(x' = r \cdot x\)
    \item \textbf{Affine masking or higher-order masking}: \(A \ne 1\), \(B \ne 0\) enables more complex transformations such as dual masking or polynomial-style encodings
\end{itemize}
Here, \(r\) represents a uniformly sampled random element drawn from the operand's algebraic domain.

\subsubsection{{\textbf{Randomness Generation}}}
The MCU interfaces with a securely seeded pseudo-random number generator (PRNG) to obtain fresh values for \(A\) and \(B\). The PRNG is implemented as a lightweight hardware module based on a Linear Feedback Shift Register (LFSR), initialized using a secure entropy source during reset. Designed for low-latency operation, it produces high-entropy random values every clock cycle, allowing the MCU to apply fresh masks to each operand without introducing pipeline stalls. Depending on the instruction’s domain:
\begin{itemize}
    \item For \(GF(2^n)\): \(A \ne 0\) is enforced to maintain multiplicative invertibility.
    \item For \(\mathbb{Z}/2^n\mathbb{Z}\): Random values are sampled modulo \(2^n\).
    \item For \(GF(2)\): Only bitwise random bits are required. \textcolor{black}{This means each operand bit is masked using a single uniformly random bit (\(r \in \{0,1\}\)), since all operations in \(GF(2)\) reduce to XOR, and no multi-bit structure or arithmetic is involved.}
\end{itemize}

For higher-order masking (\texttt{MASK\_SHARES} > 1), the MCU invokes additional PRNG rounds to produce statistically independent affine-masked shares per operand.

\subsubsection{{\textbf{Masked Operand Dispatch.}}}
Once the masking transformation is applied, the MCU outputs two masked operands corresponding to the instruction's source registers. These operands are forwarded directly to the CFU, which performs all cryptographic computations exclusively on randomized data. By ensuring that the sensitive inputs to the CFU are decorrelated from their original values, the MCU prevents data-dependent switching activity within the datapath, which is a primary contributor to power side-channel leakage.

This masking not only mitigates first-order leakage—where the power consumption of a single trace may reveal secret data—but also supports higher-order protection through the generation of multiple random shares, as configured by the \texttt{MASK\_SHARES} metadata. The CFU remains agnostic to the masking scheme: it simply processes the masked operands as it would unmasked ones, enabling seamless integration with the existing functional unit logic. This design approach achieves strong security guarantees without compromising pipeline timing or functional correctness.
\begin{figure}[t!]
    \centering
    \includegraphics[width=\linewidth]{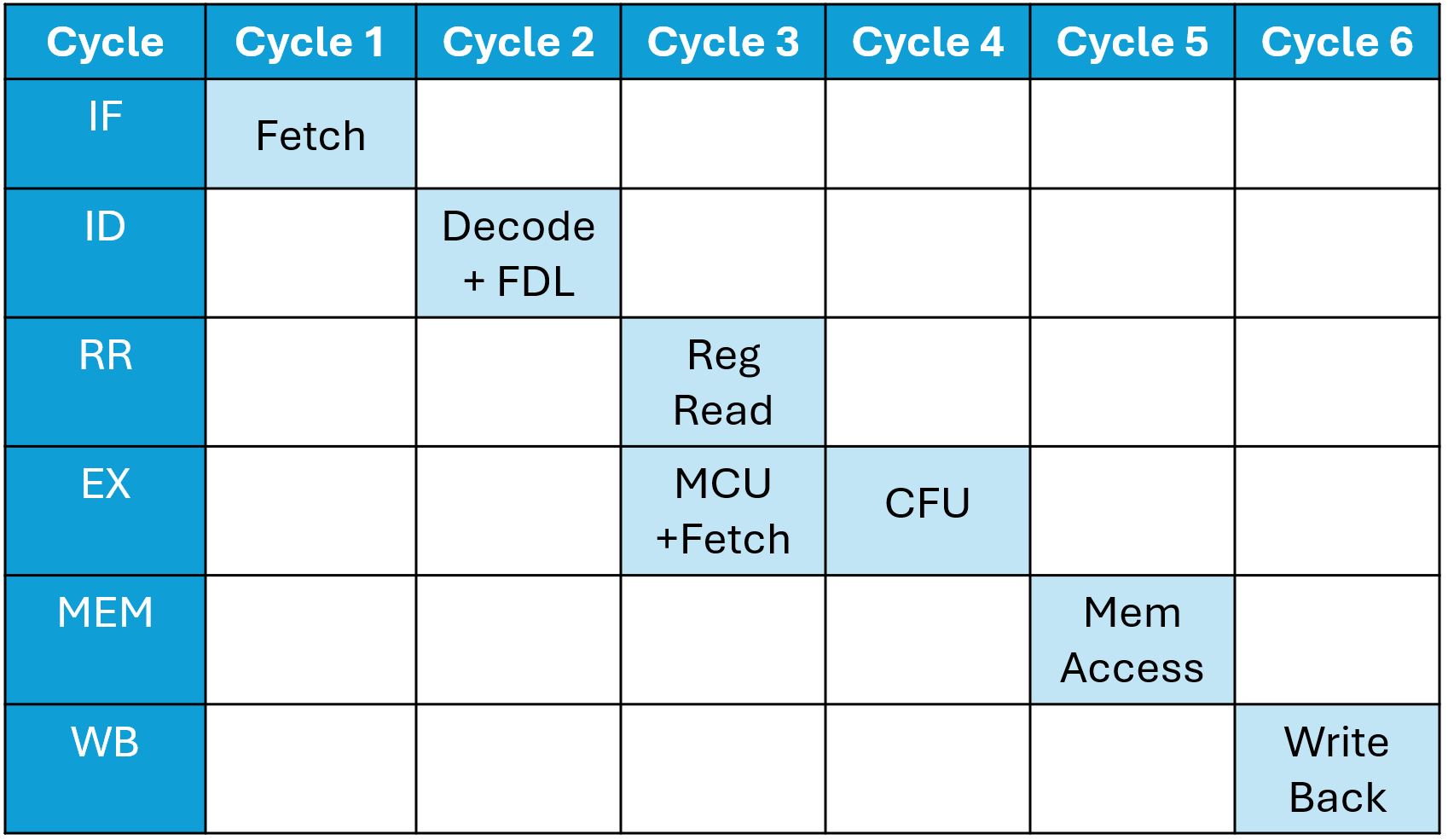}
    \caption{Timing diagram showing MCU integration within the CryptRISC pipeline. Operand masking overlaps with operand
fetch and does not delay CFU execution.}
    \label{fig:timing}
\end{figure}



\subsection{Pipeline Integration and Performance}

Figure~\ref{fig:timing} illustrates the integration of field-aware masking components into the six-stage CryptRISC pipeline. The execution of a cryptographic instruction begins in the \textit{Instruction Fetch (IF)} stage during Cycle 1. In Cycle 2, the instruction enters the \textit{Instruction Decode (ID)} stage, where it is decoded and processed by the FDL to identify the masking domain (Boolean, affine, or arithmetic).
\textcolor{black}{Although CryptRISC follows a six-stage organization, the Decode and Register Read operations are implemented within the same pipeline stage (Cycle 2) as in the original CVA6 design, using parallel datapaths.} \textcolor{black}{In Cycle 3, the instruction advances to the \textit{Execute (EX)} stage, where the Register Read and MCU logic operate concurrently. Operands are fetched from the register file at the beginning of this cycle and immediately forwarded to the MCU, which performs the masking transformation using combinational logic. This enables the MCU to complete within the same cycle without introducing additional latency or altering the pipeline schedule.}

\textcolor{black}{While we refer to a six-stage pipeline in Section 5 for conceptual clarity (treating Decode and Register Read as distinct stages), CryptRISC follows the original CVA6 microarchitecture, where Decode and Register Read occur within the same pipeline cycle using parallel datapaths. Consequently, the MCU operates entirely within the timing budget of the EX stage and was verified not to be on the processor’s critical path post-synthesis and timing analysis.}

The MCU performs operand masking within a single clock cycle, enabling secure operand transformation without introducing any pipeline stalls or structural hazards. This tight integration preserves the throughput of the pipeline and is fully compliant with the CryptRISC valid/ready handshake protocol. In Cycle 4, the CFU consumes the masked operands and performs the designated cryptographic operation. The pipeline proceeds with the \textit{Memory Access (MEM)} and \textit{Write Back (WB)} stages in Cycles 5 and 6, respectively.
Importantly, the MCU’s integration does not require any modifications to the RISC-V instruction encoding format. The masking logic is dynamically configured at runtime through metadata generated during decode, ensuring seamless execution of secure cryptographic operations with no performance degradation.

Therefore, the proposed method offers several key benefits that make it both efficient and practical for secure cryptographic execution:
\begin{itemize}
    \item \textbf{Affine masking unification}: A single datapath supports multiple masking schemes, reducing hardware duplication.
    \item \textbf{Runtime reconfigurability}: Dynamically adapts masking strategy based on field classification metadata.
    \item \textbf{Scalable to higher-order security}: Share expansion is driven by metadata without structural changes.
    \item \textbf{ISA transparency}: Operates entirely in the microarchitecture with no impact on software.
    \item \textbf{Low-latency integration}: Executes in a single pipeline cycle with minimal control overhead.
\end{itemize}


These advantages collectively establish our masking architecture as a scalable, efficient, and software-transparent defense mechanism for securing cryptographic workloads against power side-channel attacks. We now evaluate the performance, leakage characteristics and  hardware overhead of the proposed CryptRISC design in the subsequent sections.

\textcolor{black}{While the individual masking schemes are based on established cryptographic principles, the novelty of this work lies in the architectural framework that automates their selection and deployment at the instruction level.
Even for \textbf{new cryptographic algorithms} such as post-quantum encryption, CryptRISC requires only a new instruction format and field tagging in the ISA, with minimal changes to the FDL. \ul{To our knowledge, CryptRISC is the first to enable fine-grained, field-aware operand masking within a general-purpose cryptographic ISA}.} 
\section{Results}
\label{sec:results}

In this section, we present the results of our evaluation of the proposed CryptRISC processor, focusing on performance, hardware overhead, and resistance to PSC attacks.

\subsection{Experimental Setup}
To evaluate the performance and side-channel resistance of our masked CVA6 RISC-V core, we utilized the Digilent Genesys 2 FPGA development board. The Genesys 2 platform features a Xilinx Kintex-7 XC7K325T-2FFG900C FPGA, providing sufficient resources for implementing complex processor designs. This platform has been previously used in RISC-V core evaluations, making it an ideal choice for our experiments \cite{sa2023cva6}.

We employed the RISC-V Cryptography Extensions benchmarking suite, which includes widely used cryptographic algorithms, namely AES, SHA-256, SHA-512, SM3, and SM4 \cite{riscvcrypto}. These benchmarks were executed on both the baseline CVA6 core and our CryptRISC core with the RISC-V scalar cryptography extension and integrated masking framework. Execution cycles were measured using the 64-bit cycle counter available in the RISC-V architecture's CSR register, ensuring precise measurement.

\subsection{Evaluation Metrics}
We evaluated and compared the performance and side-channel resilience of the Baseline CVA6 and CryptRISC across the following metrics:

\begin{enumerate}
    \item 
    \textbf{Execution Time}: Execution time refers to the total duration required for a benchmark to complete its cryptographic operation, measured from the start of the algorithm to its completion. It directly reflects the number of cycles executed, as given by equation 1: \begin{equation} \text{Execution Time} = \frac{\text{Cycle Count}}{\text{Clock Frequency}} \end{equation} Since the clock frequency remains constant between Baseline CVA6 and CryptRISC, execution time effectively captures the impact of hardware optimizations that reduce cycle count. In addition to capturing cycle count, execution time encapsulates the cumulative impact of multiple performance factors, including memory accesses, pipeline stalls, and cache behavior. This makes it a reliable metric that not only reflects the efficiency of instruction execution, but also provides a complete view of system performance across different configurations.

    \item \textbf{Speedup}: This metric captures the relative performance improvement achieved by CryptRISC. Speedup is defined as the ratio of the execution time of the Baseline CVA6 to the execution time of CryptRISC: \begin{equation} \text{Speedup} = \frac{\text{Baseline CVA6 Time}}{\text{CryptRISC Time}} \end{equation} 

    \item \textbf{Memory Footprint:} The total size of the compiled benchmark binary (in bytes), including all segments necessary for program execution. This metric captures the static code efficiency of the design. A reduced memory footprint is especially valuable for embedded and resource-constrained systems, enabling more efficient use of on-chip memory and facilitating deployment in lightweight environments.

    \item  \textbf{Test Vector Leakage Assessment (TVLA):} To evaluate the side-channel resistance of \textbf{CryptRISC's field-aware masking} architecture, we adopt TVLA as the primary metric. TVLA is a widely used statistical method for detecting power side-channel leakage in cryptographic hardware~\cite{ding2018towards, ding2016simpler}. It involves comparing power traces collected under two input conditions—fixed and random—using Welch’s \emph{t}-test to assess statistical deviation at each sample point. Under the null hypothesis (\textit{i.e.}, no leakage), \emph{t}-values should follow a normal distribution centered at zero. A threshold of $\pm4.5$, corresponding to a false positive rate below $10^{-7}$, is commonly used to indicate statistically significant leakage~\cite{schneider2016leakage, ferrufino2023fobos}. Fewer \emph{t}-values exceeding this threshold imply stronger resistance to first-order power side-channel attacks.

\end{enumerate}

\subsection{Performance Evaluation}
To evaluate the performance improvements achieved by our CryptRISC core, we conducted a comprehensive evaluation using a set of benchmarks across the two RISC-V core configurations.


\paragraph{\textbf{Benchmark Setup}:} To comprehensively evaluate the impact of our proposed design, we utilized a diverse set of cryptographic benchmarks. The benchmark setup involved the following:

\begin{itemize} \item \textbf{Software Implementation:} The baseline software implementations used standard cryptographic libraries, such as OpenSSL compiled for the RISC-V architecture \cite{viega2002network}. These libraries provide software-based implementations of popular cryptographic algorithms optimized for generic processor architectures without hardware acceleration. The software implementations serve as a reference for evaluating the performance improvement gained by leveraging hardware-based cryptographic extensions.

\item \textbf{CryptRISC Implementation:} The benchmarks for the evaluation of CryptRISC were derived from the RISC-V Crypto Benchmark Suite, which evaluates cryptographic algorithms on RISC-V architectures. These benchmarks are meant to be executed on extended variants of the RISC-V architecture that leverage the proposed scalar cryptography extensions to accelerate cryptographic workloads.

\end{itemize}

\begin{table}[h]
    \centering
    \caption{Execution Time and Speedup Across CVA6 Configurations.}
    \label{tab:execution_time_speedup}
    \begin{tabular}{l
        >{\centering\arraybackslash}p{1.8cm}
        >{\centering\arraybackslash}p{1.8cm}
        >{\centering\arraybackslash}p{1.8cm}}
        \toprule
        \textbf{Benchmark} &
        \shortstack{\textbf{Baseline} \\ \textbf{CVA6 (ms)}} &
        \shortstack{\textbf{CryptRISC} \\ \textbf{(ms)}} &
        \textbf{Speedup} \\
        \midrule
        AES-128   & 7352.63  & 1664.79  & 4.42$\times$ \\
        AES-192   & 1961.54  & 344.03   & 5.75$\times$ \\
        AES-256   & 1498.34  & 300.06   & 5.50$\times$ \\
        SHA-256   & 3186.22  & 740.05   & 4.30$\times$ \\
        SHA-512   & 3240.69  & 1659.80  & 3.95$\times$ \\
        SM3       & 3502.87  & 514.70   & 6.80$\times$ \\
        SM4       & 1400.68  & 530.92  & 2.63$\times$ \\
        \bottomrule
    \end{tabular}
\end{table}

The performance results for the evaluated cryptographic benchmarks are summarized in Table~\ref{tab:execution_time_speedup}. The first column lists the cryptographic benchmarks, the second and third columns display the execution times in milliseconds (ms) for the Baseline CVA6 and CryptRISC, respectively. The final column shows the corresponding speedup values, which quantify the performance improvement achieved by CryptRISC. The results indicate that CryptRISC achieves substantial performance improvements across most benchmarks, with an average speedup of 4.76$\times$. SM3 demonstrates the highest speedup of 6.80$\times$, followed by AES-192 at 5.75$\times$ and AES-256 at 5.50$\times$. SHA-256 and SHA-512 also show notable speedups of 4.30$\times$ and 3.95$\times$, respectively. These results highlight the effectiveness of CryptRISC in accelerating cryptographic workloads while maintaining security through the proposed masking mechanism. 



\begin{table}[h]
    \centering
    \caption{Total Memory Footprint: Baseline CVA6 vs. CryptRISC.}
    \label{tab:memory_footprint_breakdown}
    \begin{tabular}{lccc}
        \toprule
        \textbf{Benchmark} &
        \shortstack[t]{\textbf{Baseline} \\ \textbf{CVA6 (B)}} &
        \shortstack[t]{\textbf{CryptRISC} \\ \textbf{(B)}} &
        \textbf{Reduction} \\
        \midrule
        AES-128   & 73650 & 63452 & 13.8\% \\
        AES-192   & 73736 & 63552 & 13.8\% \\
        AES-256   & 73750 & 63608 & 13.8\% \\
        SHA-256   & 67894 & 63316 & 6.7\%  \\
        SHA-512   & 69294 & 63050 & 9.0\%  \\
        SM3       & 73298 & 72912 & 0.5\%  \\
        SM4       & 62688 & 61556 & 1.8\%  \\
        \bottomrule
    \end{tabular}
\end{table}

Table~\ref{tab:memory_footprint_breakdown} presents a comparison of the total memory usage in bytes for each cryptographic benchmark across Baseline CVA6 and CryptRISC implementations. As seen in Table \ref{tab:execution_time_speedup}, the use of hardware-accelerated instructions in CryptRISC leads to consistent reductions in binary size across all evaluated algorithms. The most significant improvements are observed in AES-based workloads. AES-128, AES-192, and AES-256 each show a 13.8\% reduction in total memory usage. This is primarily due to the replacement of software-based encryption rounds and key expansion logic with single-instruction hardware primitives such as \texttt{saes64.encs}, \texttt{saes64.decsm}, and \texttt{saes64.ks1}. These specialized instructions eliminate the need for large unrolled loops and S-box handling in software. SHA-256 and SHA-512 also demonstrate significant reductions of 6.7\% and 9.0\%, respectively. In these cases, hardware acceleration streamlines logical operations and modular arithmetic into tightly encoded scalar instructions, leading to measurable code size savings. SM3 and SM4 further benefit from the hardware-backed implementation of permutation and key scheduling operations, resulting in footprint reductions of 0.5\% and 1.8\%, respectively.

These results confirm that CryptRISC not only accelerates execution but also improves memory efficiency by reducing the static binary size of cryptographic workloads. This dual benefit strengthens the practicality of deploying CryptRISC in real-world secure systems with tight memory budgets.

\begin{figure*}[t!]
    \centering

    \begin{subfigure}[b]{0.3\textwidth}
        \centering
        \includegraphics[width=\textwidth]{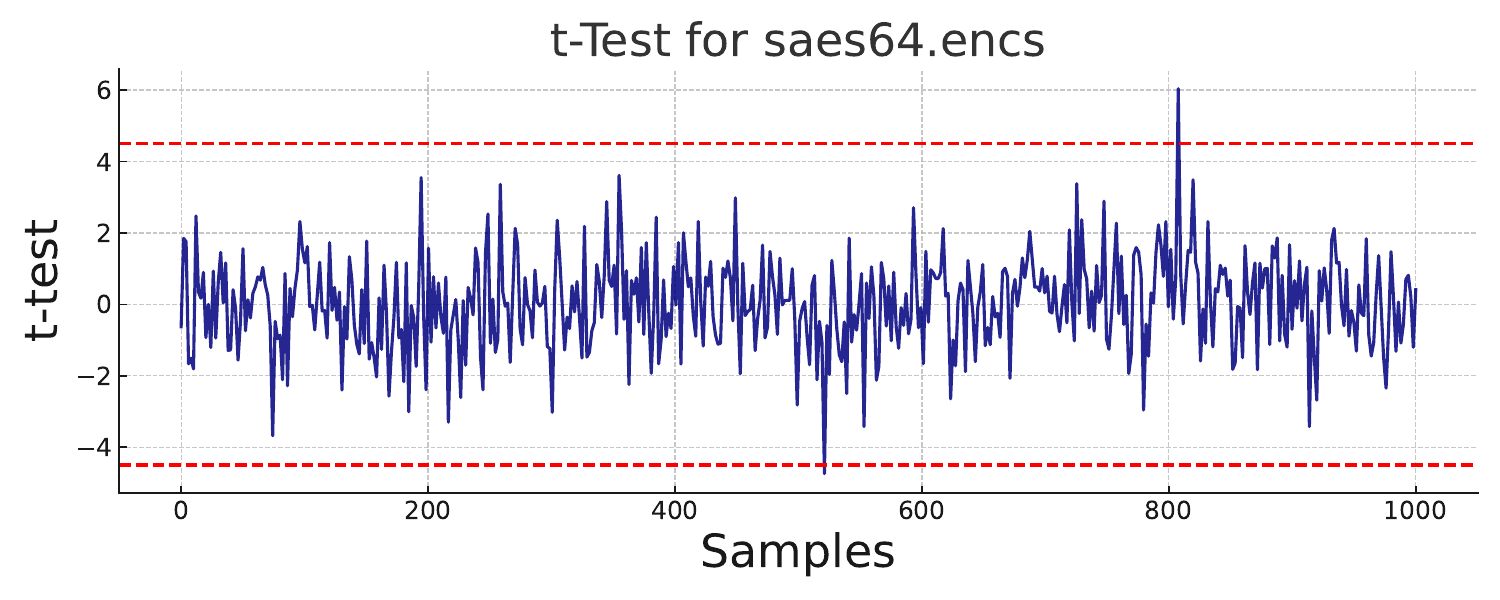}
        \caption{saes64.encs}
        \label{fig:saes64_encs}
    \end{subfigure}
    \hspace{0.03\textwidth}
    \begin{subfigure}[b]{0.3\textwidth}
        \centering
        \includegraphics[width=\textwidth]{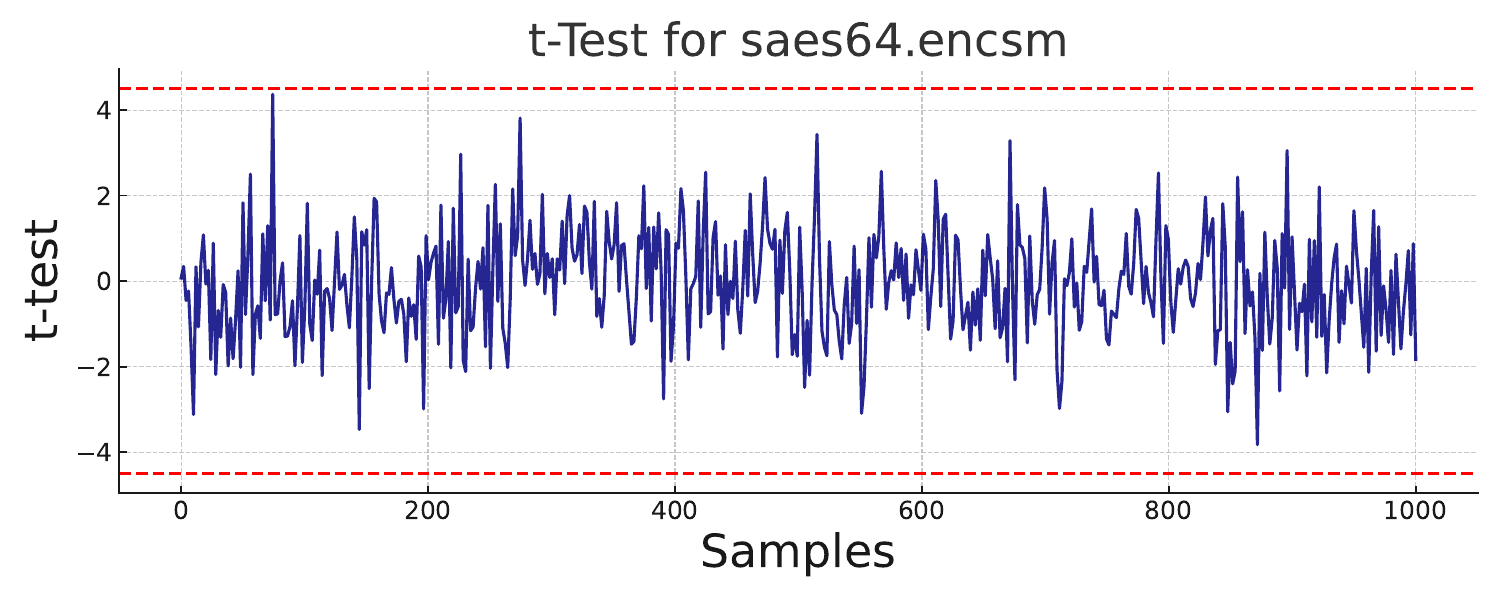}
        \caption{saes64.encsm}
        \label{fig:saes64_encsm}
    \end{subfigure}
    \hspace{0.03\textwidth}
    \begin{subfigure}[b]{0.3\textwidth}
        \centering
        \includegraphics[width=\textwidth]{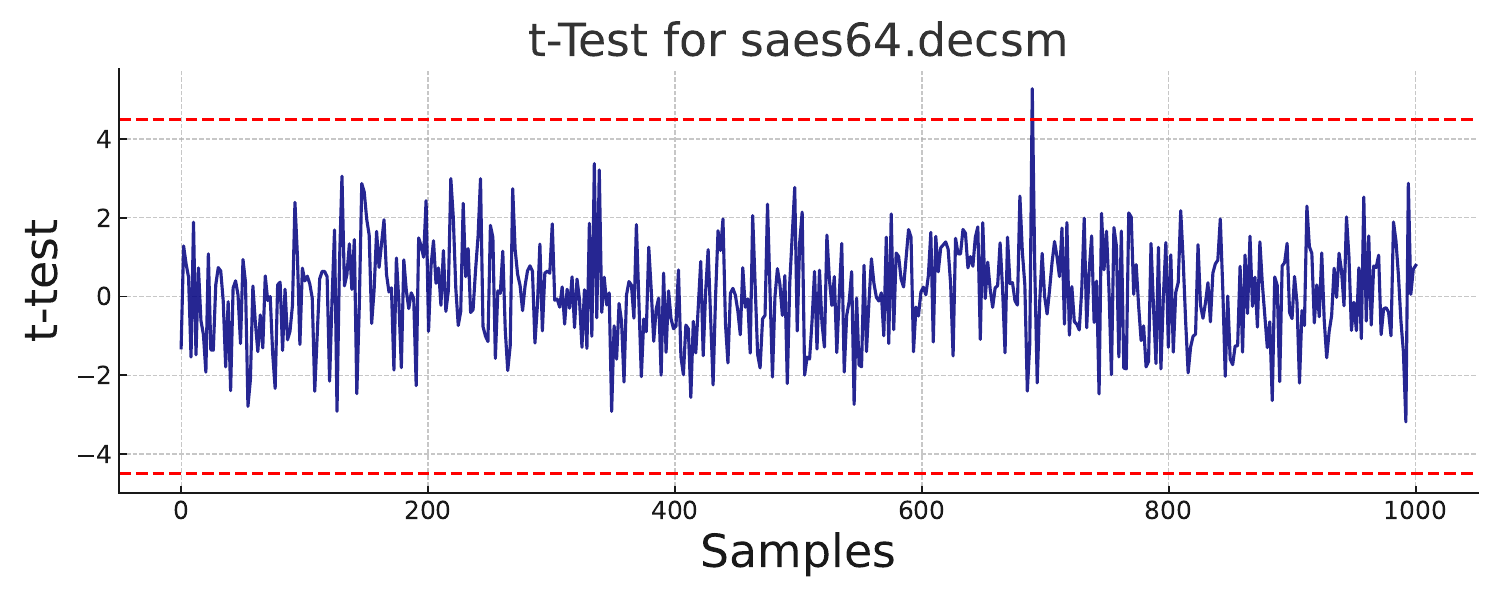}
        \caption{saes64.decsm}
        \label{fig:saes64_decsm}
    \end{subfigure}

    \begin{subfigure}[b]{0.3\textwidth}
        \centering
        \includegraphics[width=\textwidth]{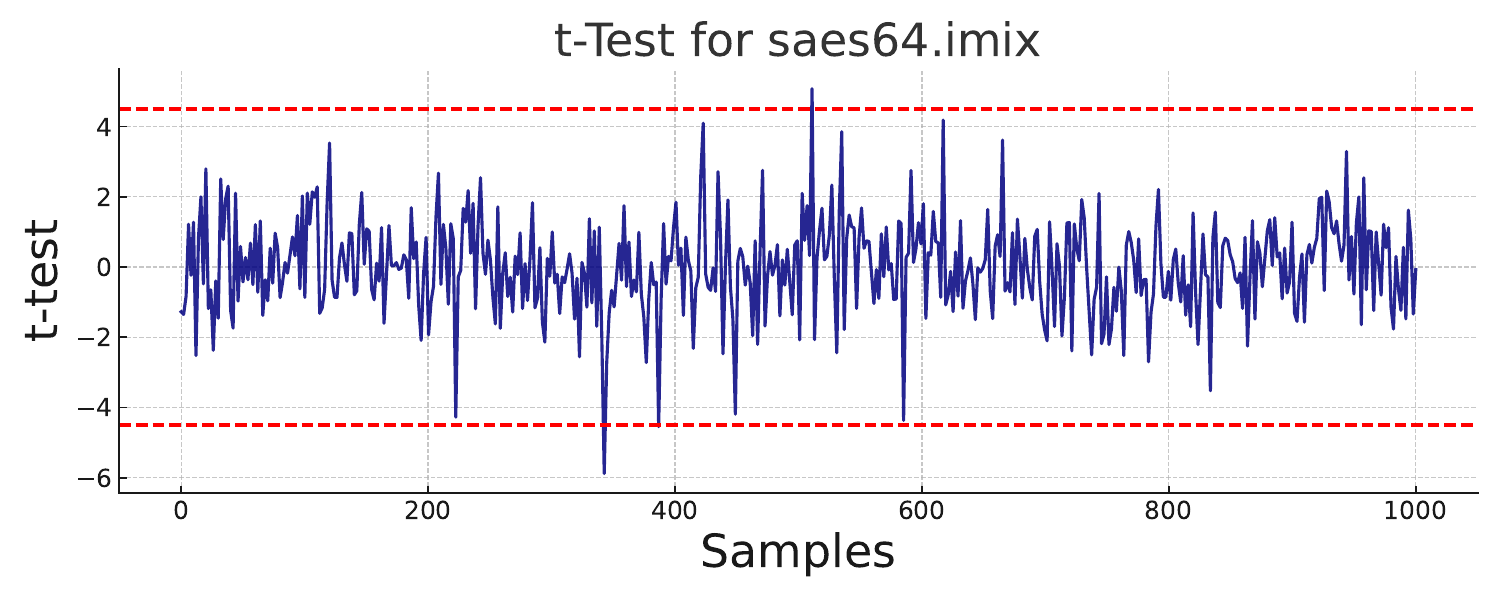}
        \caption{saes64.imix}
        \label{fig:saes64_imix}
    \end{subfigure}
    \hspace{0.03\textwidth}
    \begin{subfigure}[b]{0.3\textwidth}
        \centering
        \includegraphics[width=\textwidth]{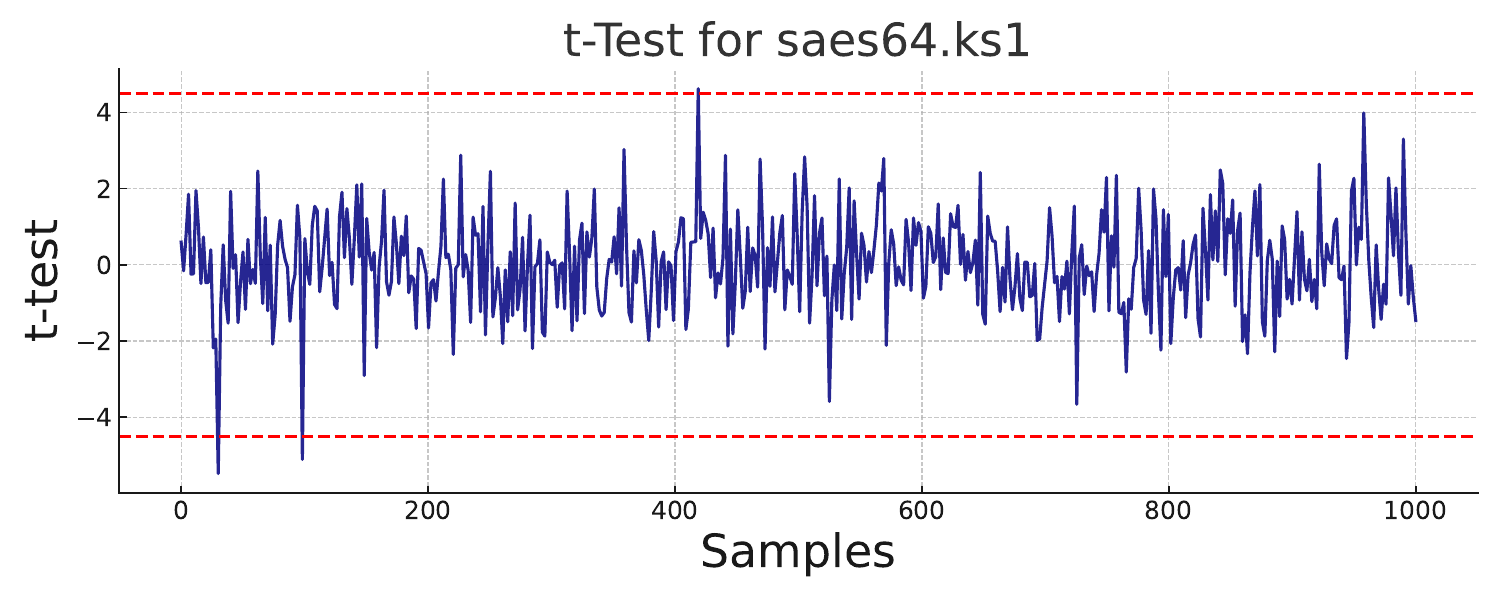}
        \caption{saes64.ks1}
        \label{fig:saes64_ks1}
    \end{subfigure}
    \hspace{0.03\textwidth}
    \begin{subfigure}[b]{0.3\textwidth}
        \centering
        \includegraphics[width=\textwidth]{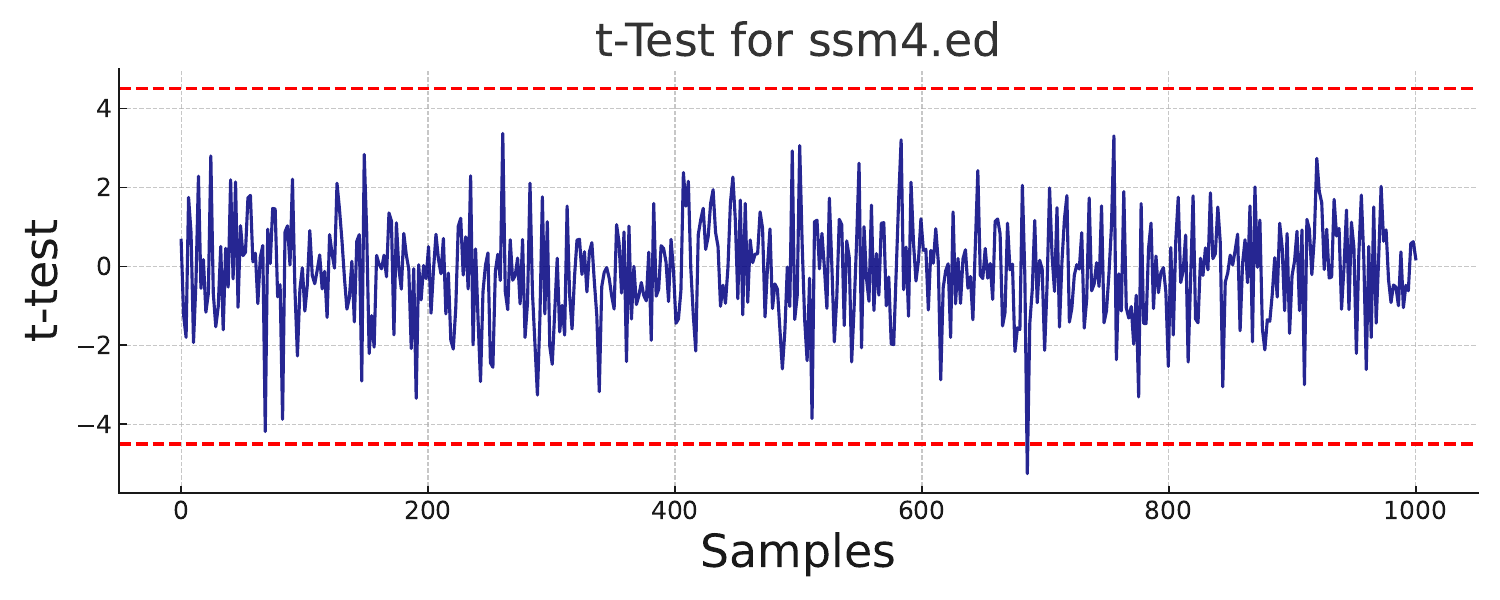}
        \caption{ssm4.ed}
        \label{fig:ssm4_ed}
    \end{subfigure}

    \begin{subfigure}[b]{0.3\textwidth}
        \centering
        \includegraphics[width=\textwidth]{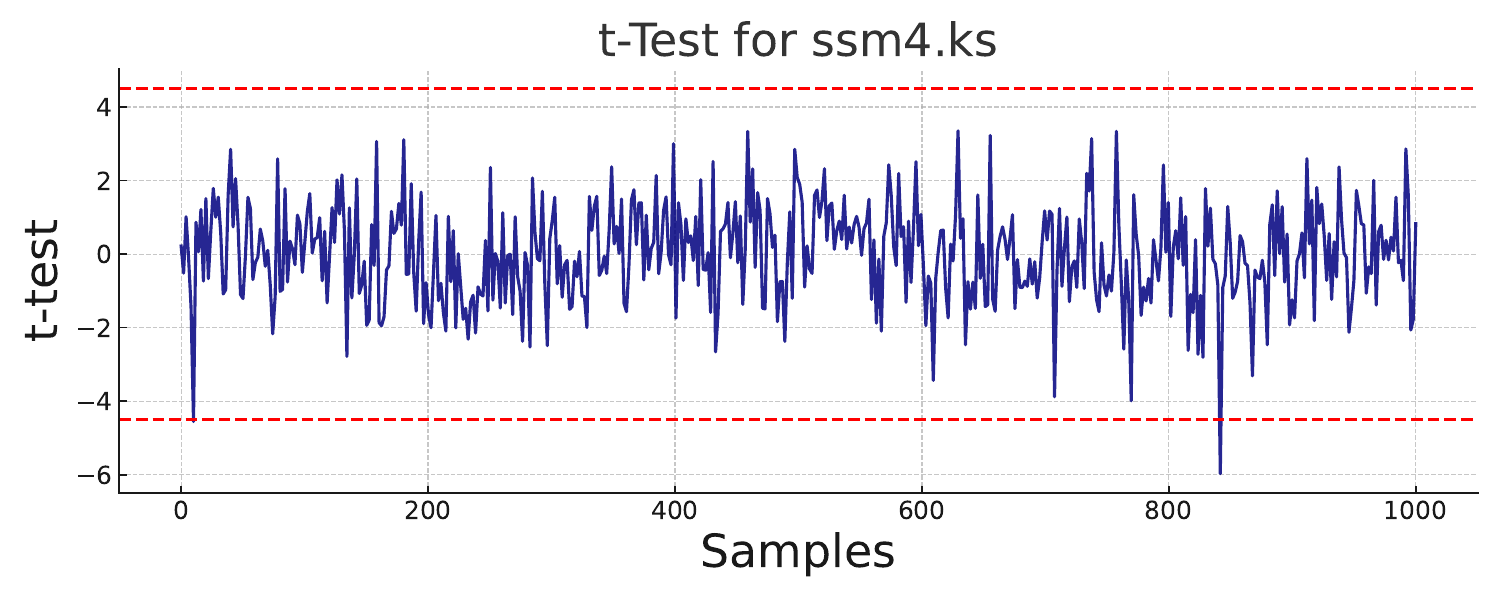}
        \caption{ssm4.ks}
        \label{fig:ssm4_ks}
    \end{subfigure}
    \hspace{0.03\textwidth}
    \begin{subfigure}[b]{0.3\textwidth}
        \centering
        \includegraphics[width=\textwidth]{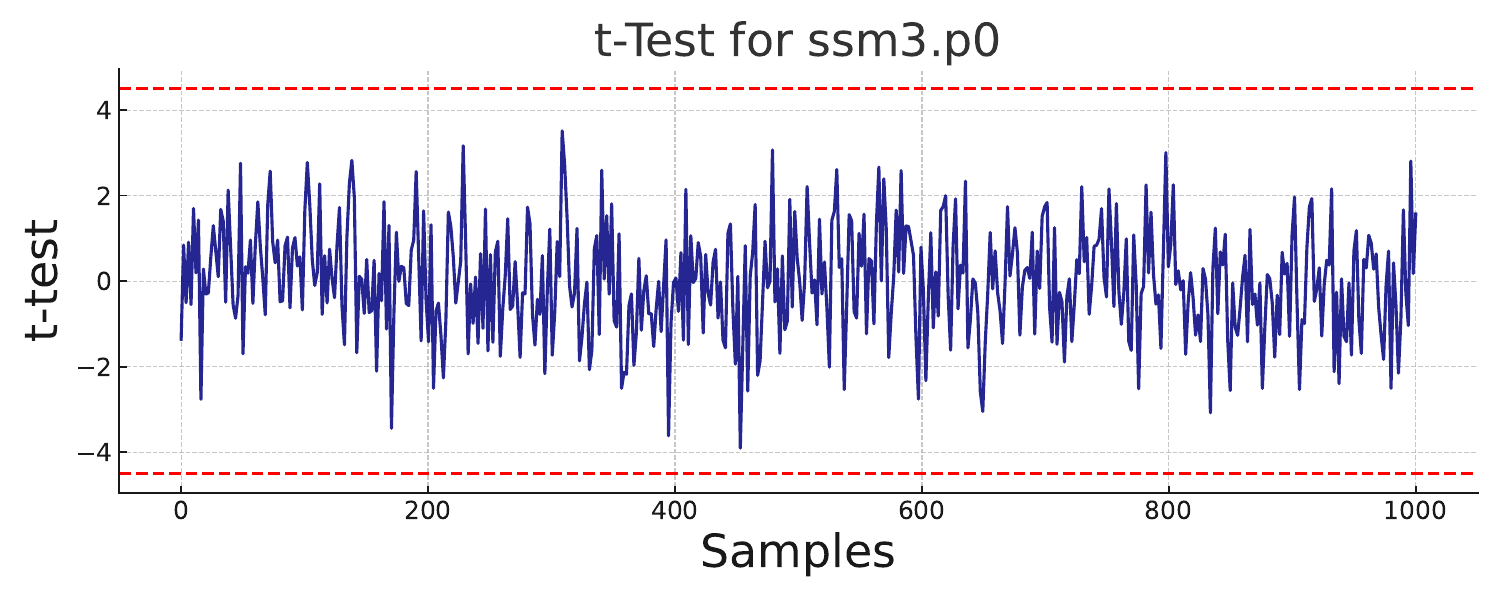}
        \caption{ssm3.p0}
        \label{fig:ssm3_p0}
    \end{subfigure}
    \hspace{0.03\textwidth}
    \begin{subfigure}[b]{0.3\textwidth}
        \centering
        \includegraphics[width=\textwidth]{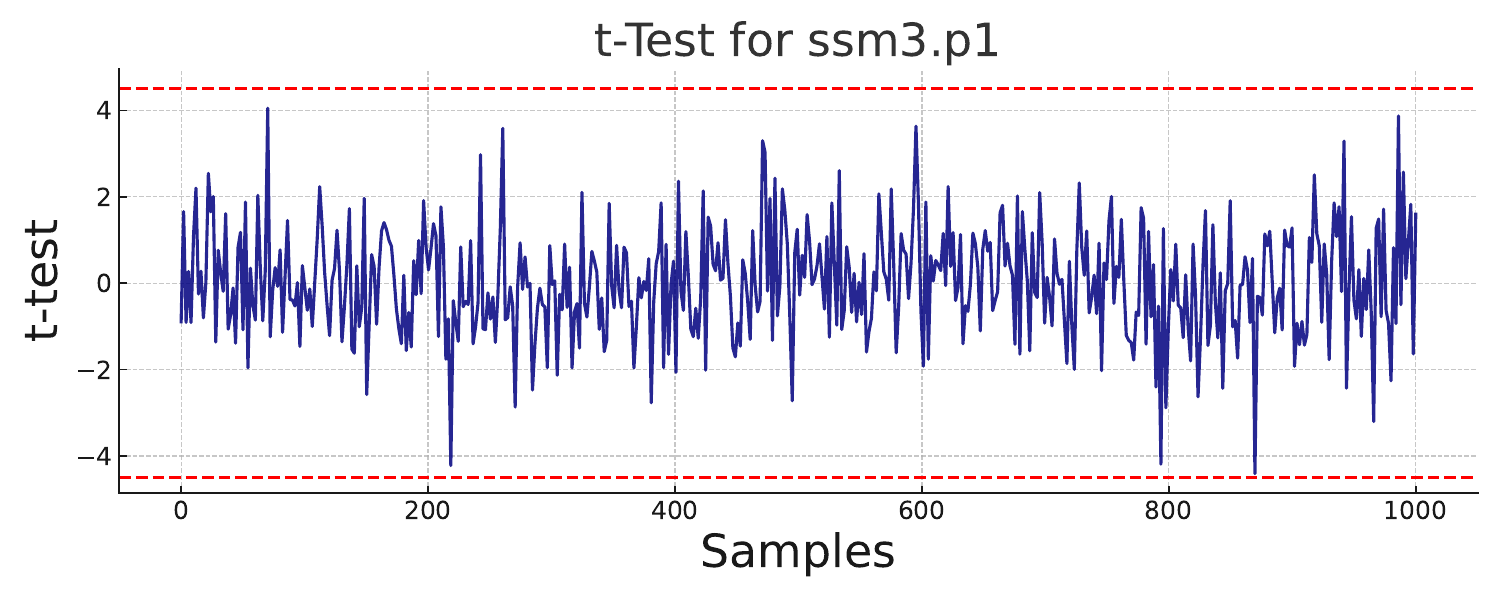}
        \caption{ssm3.p1}
        \label{fig:ssm3_p1}
    \end{subfigure}

    \caption{t-Test results for 64-bit AES, SM4, and SM3 operations in Cryptrisc . All instructions demonstrate masked behavior with limited leakage.}
    \label{fig:tvla_crypto_clean9}
\end{figure*}

\subsection{Side-Channel Resilience Results}

\begin{figure*}[t!]
    \centering

    \begin{subfigure}[b]{0.3\textwidth}
        \centering
        \includegraphics[width=\textwidth]{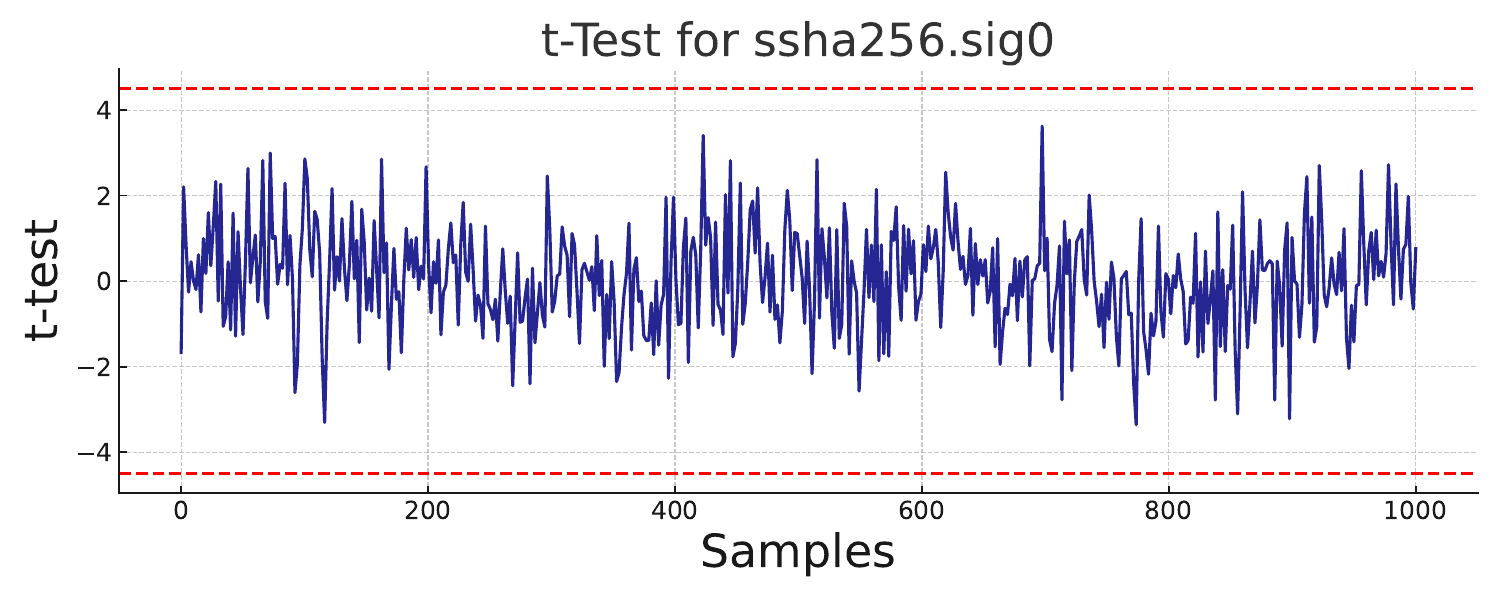}
        \caption{ssha256.sig0}
        \label{fig:ssha256_sig0}
    \end{subfigure}
    \hspace{0.03\textwidth}
    \begin{subfigure}[b]{0.3\textwidth}
        \centering
        \includegraphics[width=\textwidth]{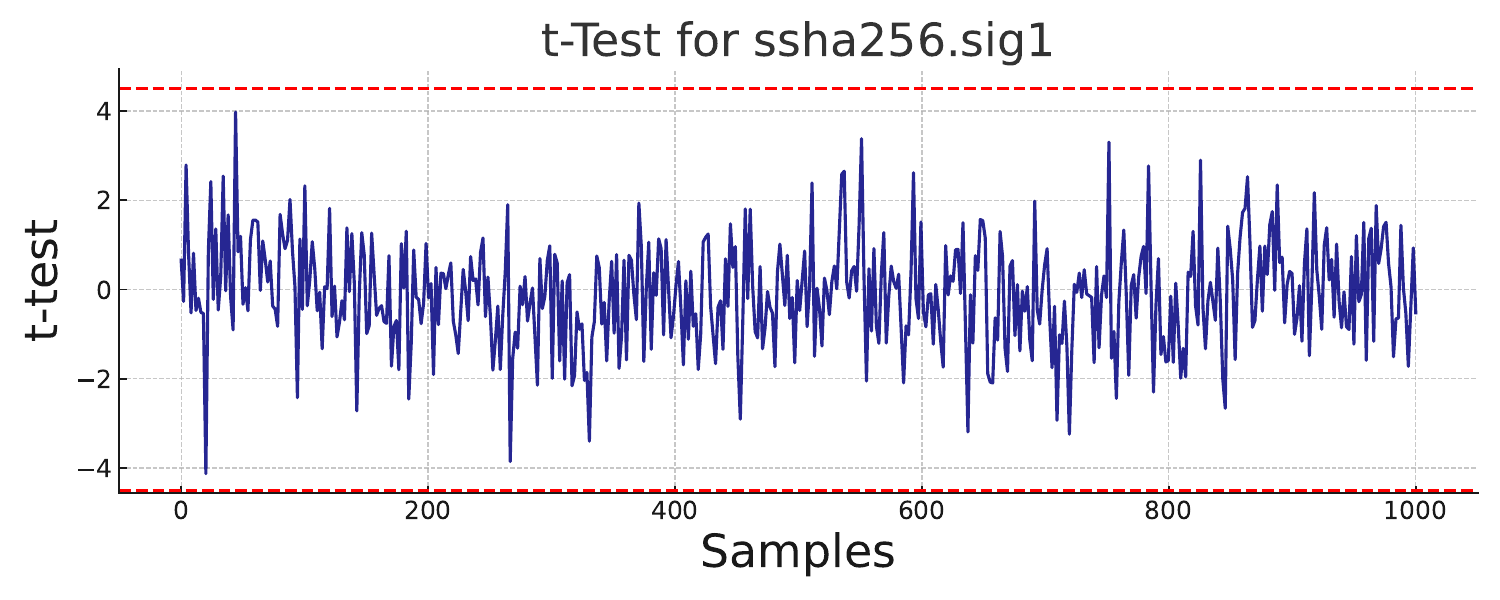}
        \caption{ssha256.sig1}
        \label{fig:ssha256_sig1}
    \end{subfigure}
    \hspace{0.03\textwidth}
    \begin{subfigure}[b]{0.3\textwidth}
        \centering
        \includegraphics[width=\textwidth]{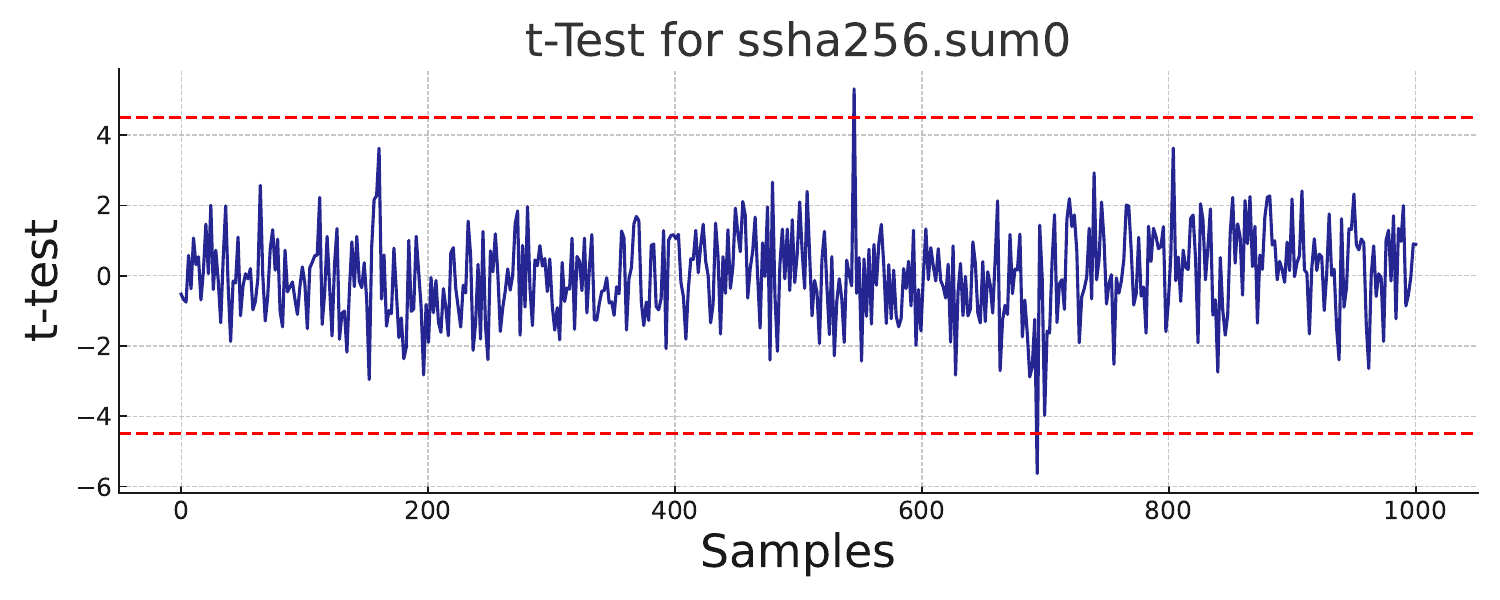}
        \caption{ssha256.sum0}
        \label{fig:ssha256_sum0}
    \end{subfigure}

    \begin{subfigure}[b]{0.3\textwidth}
        \centering
        \includegraphics[width=\textwidth]{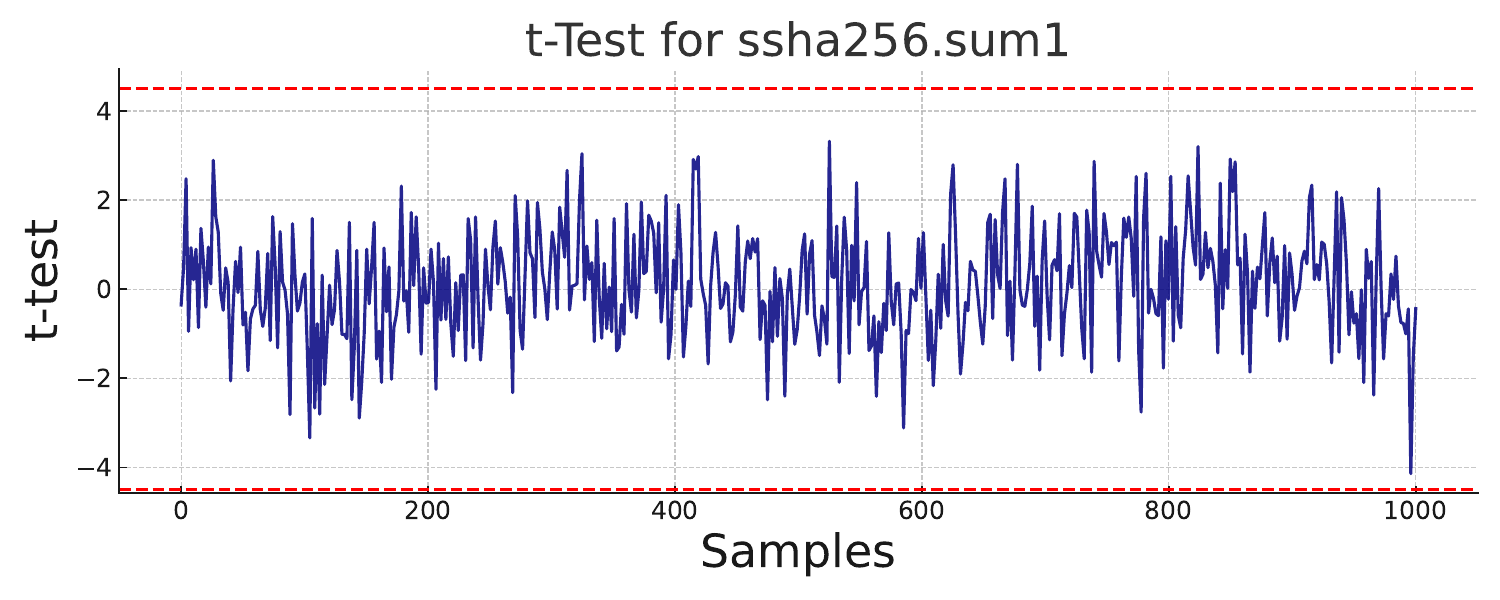}
        \caption{ssha256.sum1}
        \label{fig:ssha256_sum1}
    \end{subfigure}
    \hspace{0.03\textwidth}
    \begin{subfigure}[b]{0.3\textwidth}
        \centering
        \includegraphics[width=\textwidth]{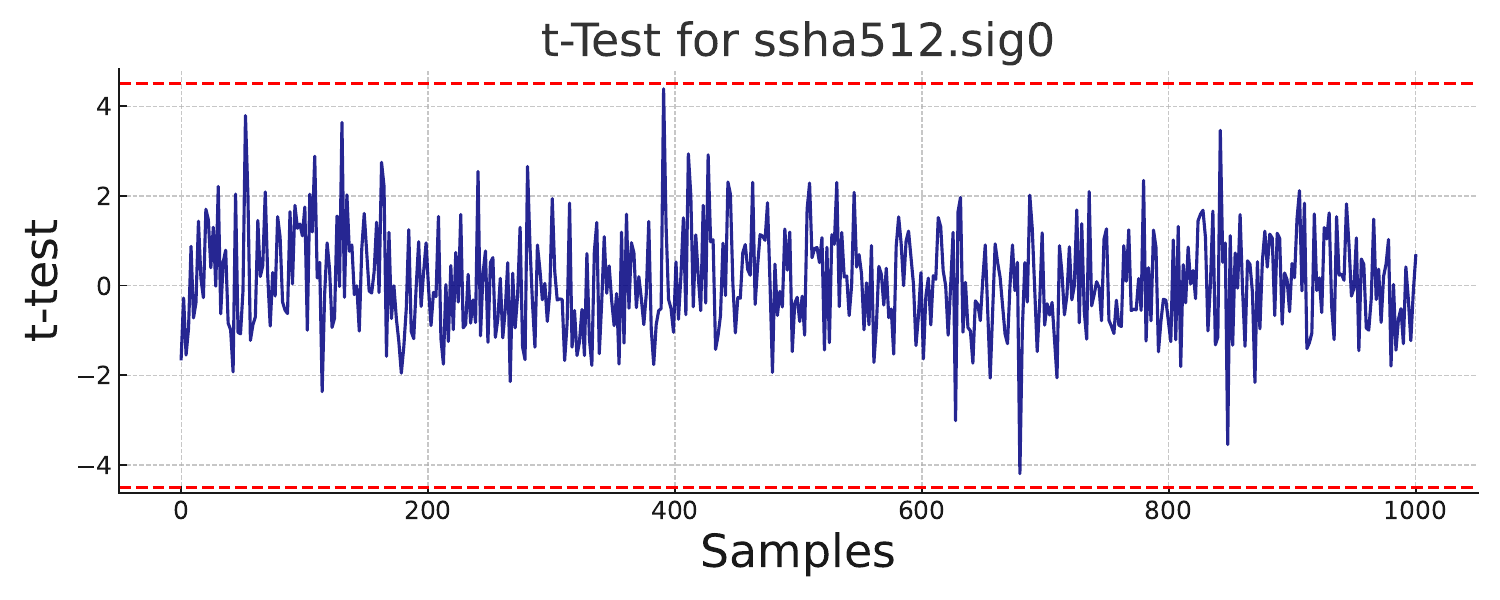}
        \caption{ssha512.sig0}
        \label{fig:ssha512_sig0}
    \end{subfigure}
    \hspace{0.03\textwidth}
    \begin{subfigure}[b]{0.3\textwidth}
        \centering
        \includegraphics[width=\textwidth]{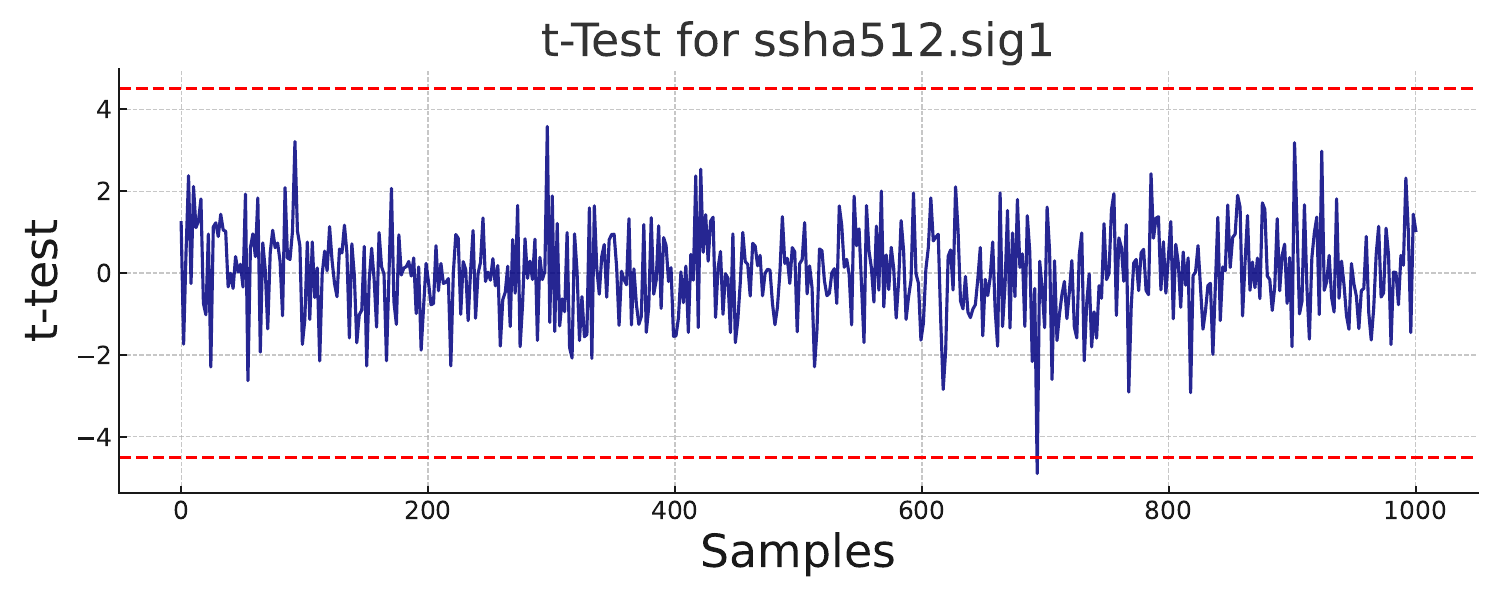}
        \caption{ssha512.sig1}
        \label{fig:ssha512_sig1}
    \end{subfigure}

    \begin{subfigure}[b]{0.3\textwidth}
        \centering
        \includegraphics[width=\textwidth]{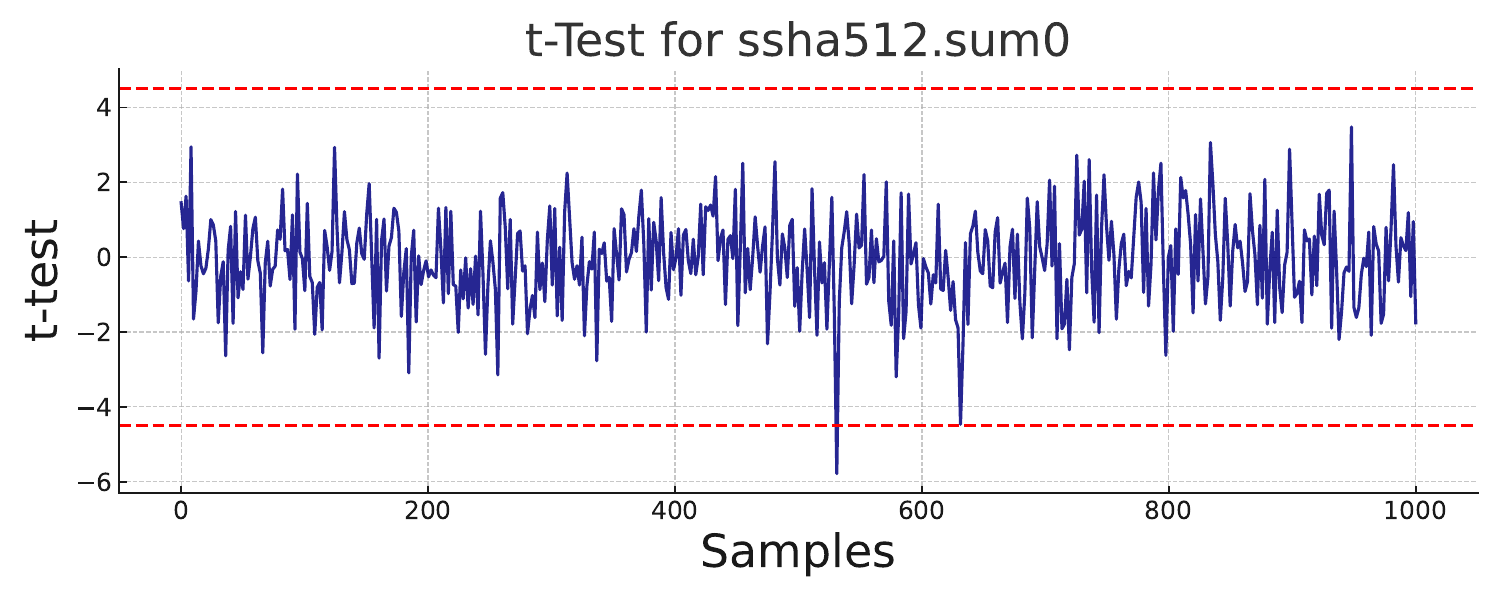}
        \caption{ssha512.sum0}
        \label{fig:ssha512_sum0}
    \end{subfigure}
    \hspace{0.03\textwidth}
    \begin{subfigure}[b]{0.3\textwidth}
        \centering
        \includegraphics[width=\textwidth]{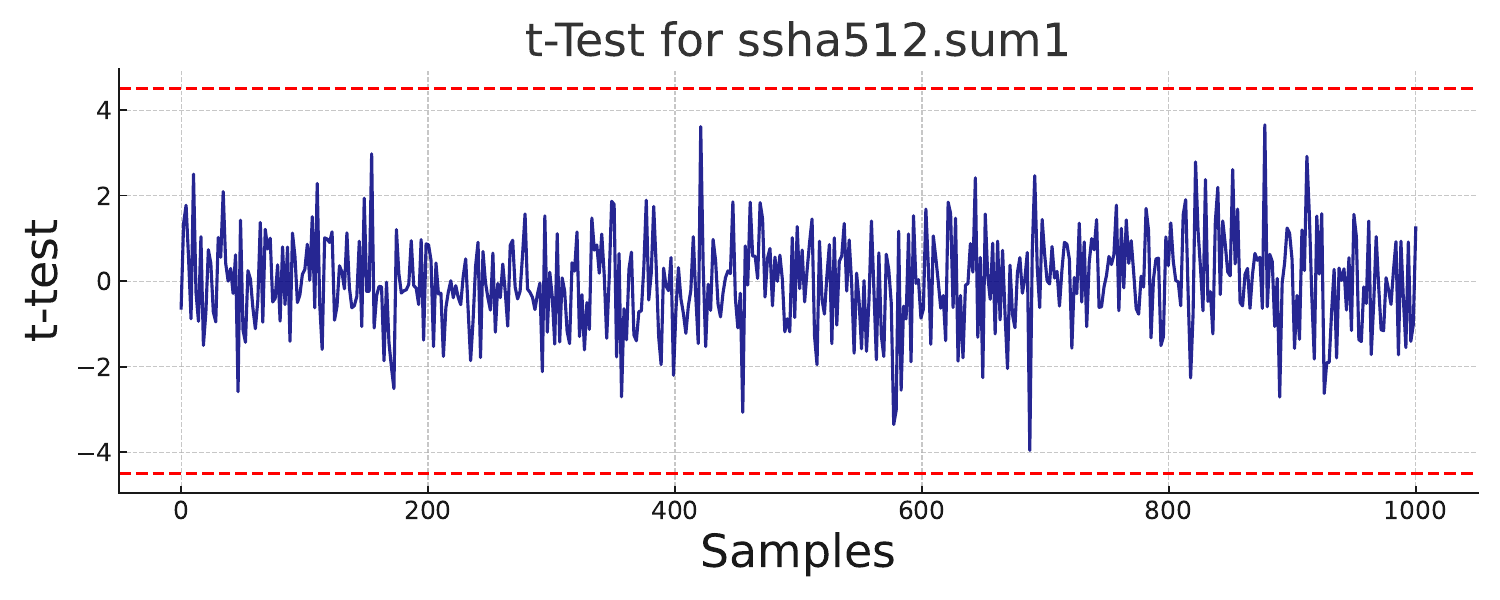}
        \caption{ssha512.sum1}
        \label{fig:ssha512_sum1}
    \end{subfigure}

    \caption{t-Test results for 64-bit SHA-2 (SHA-256, SHA-512) instructions in CryptRISC. All instructions demonstrate minimal leakage under masked conditions, with occasional expected spikes within the ±4.5 threshold.}
    \label{fig:tvla_sha_sm3_all}
\end{figure*}

To assess the effectiveness of \textbf{CryptRISC’s field-aware masking}, we conducted instruction-level TVLA across the CFU instructions. Each instruction in the CFU was executed repeatedly with fixed and random inputs, and 4000 power traces were collected. The power traces were aligned to instruction-level execution points and evaluated independently. Figures~\ref{fig:tvla_crypto_clean9} and~\ref{fig:tvla_sha_sm3_all} present the resulting \emph{t}-test values over time for each instruction. The red dashed line indicates the leakage threshold. The instructions were grouped based on their respective cryptographic algorithms and evaluated for leakage behavior in relation to their underlying operations and applied masking strategies. Next, we summarize the key observations drawn from the TVLA plots for each class of instruction:

\begin{itemize}
    \item \textbf{AES and SM4:} Instructions such as \textit{saes64.encs}, \textit{saes64.encsm}, \textit{saes64.decsm}, \textit{saes64.imix}, and \textit{ssm4.ks} show minimal statistical variance across all time samples, as observed in Figures \ref{fig:tvla_crypto_clean9}a to \ref{fig:tvla_crypto_clean9}g. The \emph{t}-value traces remain consistently flat, with values tightly\textbf{ bounded below $\pm$1.0}, and no observable excursions toward the leakage threshold. These instructions typically involve nonlinear S-box substitutions and MixColumns transformations, which are known to be vulnerable to data-dependent switching. CryptRISC applies affine masking over $GF(2^8)$ in these cases, ensuring that all intermediate values are randomized before entering the functional unit. The results confirm that the masking effectively eliminates detectable leakage, even in operations with high switching complexity.

    \item \textbf{SHA-256 and SHA-512:} The SHA-family instructions evaluated involve bitwise logical operations and modular additions. These exhibited slightly more variance than AES, as expected from their computational structure, but their \emph{t}-values remained securely below the $\pm4.5$ threshold, with most staying \textbf{within the $\pm$2 range}, as seen in Figures \ref{fig:tvla_sha_sm3_all}a to \ref{fig:tvla_sha_sm3_all}h. Boolean and arithmetic masking were applied at decode, depending on the operation type. These masking techniques successfully suppressed observable leakage despite the deterministic patterns inherent to rotations, shifts, and additions that lead to data-dependent power variation.

    \item \textbf{SM3:} The permutation-heavy SM3 instructions \textit{ssm3.p0} and \textit{ssm3.p1} displayed the highest variance among the evaluated instructions, though still within secure bounds as observed in Figures \ref{fig:tvla_crypto_clean9}h and \ref{fig:tvla_crypto_clean9}i. Minor spikes in \emph{t}-value appeared due to the bit-manipulation nature of the operations, which create scattered switching activity across the datapath. However, even in these cases, the \emph{t}-values \textbf{remained below $\pm$2.0}, well under the leakage threshold. These results indicate that field-agnostic operand masking when applied uniformly, can be effective even for instructions with irregular or high-fanout switching patterns.
\end{itemize}

Thus, across all evaluated instructions, CryptRISC maintains \emph{t}-values well below the leakage threshold of $\pm4.5$, with the majority falling within $\pm2$. These results confirm that the field-aware operand masking architecture provides robust protection against first-order power side-channel leakage when the instructions are executed in isolation.



\subsection{Hardware Overhead}

In order to assess the hardware overhead introduced by integrating the cryptographic ISE and the proposed masking mechanism into the CryptRISC core, we synthesized and analyzed resource utilization under the Baseline CVA6 and CryptRISC configurations. The evaluation focused on critical hardware metrics, including the number of Lookup Tables (LUTs) and Flip-Flops (FFs), as well as timing characteristics such as critical path delay and timing slack.

\begin{table}[h]
\centering
\caption{FPGA Resource Utilization and Timing.}
\label{tab:fpga_resource_utilization}
\begin{tabular}{p{2.8cm} c c c}
\hline
\textbf{Metric} & 

\shortstack[t]{\textbf{Baseline} \\ \textbf{CVA6}} &

\textbf{CryptRISC} & \shortstack[t]{\textbf{\%} \\ \textbf{Increment}} \\
\hline
\textbf{LUTs} & 72,462 & 73,809 & 1.86\% \\
\textbf{Flip Flops (FFs)} & 45,444 & 45,555 & 0.24\% \\
\textbf{Critical Path Delay (ns)} & 1.183 & 1.183 & 0.00\% \\
\textbf{Timing Slack (ns)} & 0.177 & 0.177 & 0.00\% \\
\textbf{Power (W)} & 2.078 & 2.115 & 1.78\% \\
\hline
\end{tabular}
\end{table}

Table~\ref{tab:fpga_resource_utilization} presents a detailed comparison of FPGA resource utilization and timing characteristics between the Baseline CVA6 and CryptRISC configurations. The table highlights the impact of adding cryptographic modules and masking logic on hardware resources and timing parameters. CryptRISC introduces a modest increase in hardware resource usage over the Baseline CVA6 core. The number of LUTs increases by only 1.86\% from 72,462 to 73,809. This is primarily due to the integration of cryptographic instruction support and operand masking logic. The number of Flip-Flops (FFs) also rises marginally by 0.24\% (from 45,444 to 45,555), reflecting the addition of control and pipeline registers needed for secure execution. Despite these additions, CryptRISC maintains identical timing characteristics compared to Baseline CVA6. Importantly, the critical path delay and timing slack remain unchanged, confirming that the integration of side-channel countermeasures does not introduce new timing bottlenecks or compromise pipeline performance.  In terms of power consumption, CryptRISC exhibits a minimal increase of 1.78\%, suggesting that the masking and cryptographic extensions have a negligible impact on overall dynamic power.

To quantify the hardware resource footprint of the cryptographic and masking logic integrated into CryptRISC, Table~\ref{tab:crypto_and_masking_lut_breakdown} presents the slice LUT usage for each individual module. The CFU, responsible for executing all scalar cryptographic instructions, is composed of dedicated submodules for the cryptographic operations. Complementing the CFU, the masking pipeline implemented across the decode and execute stages consists of the FDL and MCU, which together add only 256 LUTs to the design.


\begin{table}[h]
    \centering
    \caption{LUT Usage Breakdown for the Cryptographic and Masking Submodules.}
    \label{tab:crypto_and_masking_lut_breakdown}
    \begin{tabular}{l l c}
        \toprule
        \textbf{Component} & \textbf{Hardware Unit} & \textbf{Slice LUTs} \\
        \midrule
        AES                    & CFU & 1284 \\
        SHA-512                & CFU                           & 250  \\
        SHA-256                & CFU                           & 125  \\
        SM4                    & CFU                           & 92   \\
        SM3                    & CFU                           & 64   \\
        Operand Masking        & FDL + MCU                     & 256  \\
        \midrule
        \textbf{Total}         &                              & \textbf{2071} \\
        \bottomrule
    \end{tabular}
\end{table}

Among the cryptographic modules, the AES unit accounts for the largest share of area, consuming 1284 LUTs. The SHA-512 and SHA-256 units require 250 and 125 LUTs, respectively, while the SM3 and SM4 units have comparatively smaller footprints. The masking logic is both lightweight and modular, contributing just 256 LUTs without impacting the instruction encoding or control flow. Overall, the combined cryptographic and masking datapath occupies only 2071 LUTs.  It is important to note that the slice LUT counts reported in Table~\ref{tab:crypto_and_masking_lut_breakdown} are based on standalone synthesis of individual modules. While these figures provide a comparative understanding of each module’s relative complexity, they do not represent the actual resource utilization after integration. Final post-integration values may reduce significantly due to optimizations performed during full-system synthesis.  Overall, this breakdown confirms that CryptRISC’s side-channel resilience and cryptographic acceleration can be achieved with minimal area overhead.

\begingroup
\color{black}

\subsection{CryptRISC vs Unmasked CVA6 with CISE}
\label{sec:tvla_masked_unmasked}

\subsubsection{Performance Comparison}


\begin{table}[h]
    \centering
    \caption{\textcolor{black}{Execution Time and Overhead.}}
    \label{tab:execution_time_masking}
    \begin{tabular}{l
        >{\centering\arraybackslash}p{1.8cm}
        >{\centering\arraybackslash}p{1.8cm}
        >{\centering\arraybackslash}p{1.8cm}}
        \toprule
        \textbf{\textcolor{black}{Benchmark}} &
        \shortstack[t]{\textbf{\textcolor{black}{CVA6 +}} \\ \textbf{\textcolor{black}{Crypto ISE (ms)}}} &
        \shortstack{\textbf{\textcolor{black}{CryptRISC}} \\ \textbf{\textcolor{black}{(ms)}}} &
        \textbf{\textcolor{black}{Overhead}} \\
        \midrule
        \textcolor{black}{AES-128}   & \textcolor{black}{1513.40}  & \textcolor{black}{1664.79}  & \textcolor{black}{10.0\%} \\
        \textcolor{black}{AES-192}   & \textcolor{black}{312.80}   & \textcolor{black}{344.03}   & \textcolor{black}{10.0\%} \\
        \textcolor{black}{AES-256}   & \textcolor{black}{272.75}   & \textcolor{black}{300.06}   & \textcolor{black}{10.9\%} \\
        \textcolor{black}{SHA-256}   & \textcolor{black}{672.80}   & \textcolor{black}{740.05}   & \textcolor{black}{11.8\%} \\
        \textcolor{black}{SHA-512}   & \textcolor{black}{1509.00}  & \textcolor{black}{1659.80}  & \textcolor{black}{12.7\%} \\
        \textcolor{black}{SM3}       & \textcolor{black}{468.80}   & \textcolor{black}{514.70}   & \textcolor{black}{9.8\%} \\
        \textcolor{black}{SM4}       & \textcolor{black}{481.70}   & \textcolor{black}{530.92}   & \textcolor{black}{11.0\%} \\
        \bottomrule
    \end{tabular}
\end{table}

The performance results comparing the unmasked and masked (CryptRISC) CVA6 configurations with cryptographic instruction set extensions are presented in Table~\ref{tab:execution_time_masking}. The results indicate that CryptRISC incurs a minimal performance penalty of approximately 10–13\% across all benchmarks when compared to its unmasked counterpart. This overhead is attributed to the field-aware operand masking integrated into the cryptographic datapath.

\subsubsection{Comparison with Unmasked Implementation}
\label{sec:unmasked_cise_comparison}

To evaluate CryptRISC's operand masking, we compare its side-channel leakage against unmasked ISEs on the SCARV processor~\cite{jayasena2025ciseleaks}, which also builds on the CVA6 core—ensuring a fair microarchitectural baseline. Table~\ref{tab:tvla_comparison_expanded} shows results across AES, SHA-256, and SHA-512 scalar instructions. The SCARV processor shows significant first-order leakage, with AES operations yielding CPA \emph{p}-values below $10^{-5}$ and \emph{t}-values exceeding the $\pm4.5$ TVLA threshold. A low \emph{p}-value indicates that the observed correlation is unlikely due to chance, while values above 0.05 suggest no leakage. In contrast, CryptRISC maintains all \emph{p}-values above 0.05 across all variants, indicating statistically non-leaking behavior.

\begin{table}[h]
\centering
\caption{\textcolor{black}{Comparison of Side-Channel Leakage in AES and SHA Modules: SCARV \cite{jayasena2025ciseleaks} (Unmasked CVA6 + Crypto ISE) vs. CryptRISC (Masked)}}
\begin{tabular}{|p{1.2cm}|c|c|c|c|}
\hline
\textbf{\textcolor{black}{Work}} & \textbf{\textcolor{black}{Module}} & \shortstack[t]{\textbf{\textcolor{black}{Minimum}} \\ \textbf{\textcolor{black}{traces}}} & \shortstack[t]{\textbf{\textcolor{black}{CPA}} \\ \textbf{\textcolor{black}{$p$-value}}} & \shortstack[t]{\textbf{\textcolor{black}{Leakdown}} \\ \textbf{\textcolor{black}{Test}}} \\
\hline

\multirow{3}{=}{\textcolor{black}{SCARV \cite{jayasena2025ciseleaks}}} 
  & \textcolor{black}{SAES64}       & \textcolor{black}{5209} & \textcolor{black}{$5.47 \times 10^{-7}$} & \cellcolor{red!25}\textbf{\textcolor{black}{Fail}} \\
  & \textcolor{black}{SSHA256}      & \textcolor{black}{4896} & \textcolor{black}{$6.53 \times 10^{-7}$} & \cellcolor{red!25}\textbf{\textcolor{black}{Fail}} \\
  & \textcolor{black}{SSHA512}      & \textcolor{black}{5103} & \textcolor{black}{$7.54 \times 10^{-7}$} & \cellcolor{red!25}\textbf{\textcolor{black}{Fail}} \\
\hline
\multirow{3}{=}{\textcolor{black}{CryptRISC}} 
  & \textcolor{black}{SAES64}       & \textcolor{black}{4000} & \textcolor{black}{$>$ 0.05}              & \cellcolor{green!20}\textbf{\textcolor{black}{Pass}} \\
  & \textcolor{black}{SSHA256}      & \textcolor{black}{4000} & \textcolor{black}{$>$ 0.05}              & \cellcolor{green!20}\textbf{\textcolor{black}{Pass}} \\
  & \textcolor{black}{SSHA512}      & \textcolor{black}{4000} & \textcolor{black}{$>$ 0.05}              & \cellcolor{green!20}\textbf{\textcolor{black}{Pass}} \\
\hline
\end{tabular}
\label{tab:tvla_comparison_expanded}
\end{table}

\section{Discussion}
\label{sec:discussion}



Table~\ref{tab:countermeasure_comparison} compares CryptRISC with software-based masking~\cite{oswald2005efficient, gao2021instruction} and fixed-function hardware masking~\cite{sasdrich2020low, rivain2010provably} across common side-channel security metrics and design attributes. CryptRISC achieves stronger leakage resilience, with all \emph{t}-values bounded within $\pm2$ and CPA \emph{p}-values above 0.05. In contrast, existing defenses typically report \emph{t}-values approaching the TVLA threshold of $\pm4.5$, indicating that CryptRISC provides a tighter statistical bound on non-leakage, offering a \textbf{stronger security guarantee}. The enhanced performance of CryptRISC can be attributed to the following factors:
\begin{itemize}
    \item Unlike fixed-function masking, which applies a static scheme, CryptRISC dynamically selects Boolean, affine, or arithmetic masking based on operand semantics \cite{sasdrich2020low, rivain2010provably, fritzmann2022masked}, preventing structure-specific leakage and adapting without requiring co-processor redesign.
    \item  While software-based masking schemes like~\cite{gao2021instruction} propose ISEs to speed up software masking, they still require manual instrumentation and handwritten assembly. CryptRISC eliminates this need by making masking ISA-transparent, supporting compiler compatibility and general-purpose usage.
    \item Dedicated hardware co-processors may achieve marginally better raw performance but lack flexibility and incur per-algorithm integration costs. CryptRISC amortizes this overhead through a shared Crypto Functional Unit that supports multiple masked instructions with no re-synthesis.

    \item Complementary logic styles and instruction reordering~\cite{Chen2025paradise} offer promising protection but typically require major backend redesigns, dynamic analysis, or result in high overheads, limiting their practicality for general-purpose processors.

\end{itemize}

\begin{table}[h]
\centering
\caption{\textcolor{black}{Comparison of Countermeasures.}}
\begin{tabular}{|p{3.5cm}|c|c|}
\hline
\textbf{\textcolor{black}{Method}} & \textbf{\textcolor{black}{\emph{t}-value}} & \textbf{\textcolor{black}{CPA \emph{p}-value}} \\
\hline
\textcolor{black}{Software-based Masking~\cite{oswald2005efficient, gao2021instruction}} & \textcolor{black}{$< \pm4.5$ } & \textcolor{black}{$> 0.05$ } \\
\hline
\textcolor{black}{Fixed-function Hardware Masking~\cite{sasdrich2020low, rivain2010provably}} & \textcolor{black}{$< \pm4.5$} & \textcolor{black}{$> 0.05$} \\
\hline
\textbf{\textcolor{black}{CryptRISC}} & \textbf{\textcolor{black}{$< \pm2$}} & \textbf{\textcolor{black}{$> 0.05$}} \\
\hline
\end{tabular}
\label{tab:countermeasure_comparison}
\end{table}

\endgroup
\section{Conclusion}
\label{sec:conclusion}

In this paper, we presented, CryptRISC, the first RISC-V-based processor integrating cryptographic acceleration with hardware-level power side-channel resistance through an ISA-driven operand masking framework. The open-source CVA6 core is extended with 64-bit RISC-V Scalar Cryptography Extensions, enabling a flexible and efficient hardware-level mechanism for dynamic, field-specific masking. This allows the processor to select appropriate masking schemes based on the underlying algebraic field of each cryptographic instruction, ensuring optimized protection against power side-channel vulnerabilities. Instruction-level Test Vector Leakage Assessment (TVLA) confirms the robustness of our masking architecture: all observed \emph{t}-values remain well below the leakage threshold of $\pm4.5$, with the majority falling within $\pm2$, indicating the absence of statistically significant first-order leakage. In addition to strong side-channel resistance, CryptRISC significantly improves cryptographic performance, achieving speedups of up to 6.80$\times$ across a wide range of commonly used cryptographic algorithms compared to baseline software implementations. By embedding the masking mechanism directly into the execution pipeline, we offer an efficient and scalable solution that avoids the complexities and inefficiencies of software-based or fixed-function hardware masking. Notably, this enhanced security is achieved with a total hardware overhead of only 1.86\%, demonstrating CryptRISC's practicality for secure embedded systems.

\balance
\bibliographystyle{ACM-Reference-Format}
\bibliography{sample-base}

\end{document}